\documentclass[nofootinbib,twocolumn,showpacs,preprintnumbers,amsmath,amssymb,floatfix]{revtex4}
\usepackage{graphicx}
\usepackage{dcolumn}
\usepackage{bm}
\newcommand{\Nsp}  	{N_{\mathrm{sp}}}
\newcommand{\Nmax}  	{N_{\mathrm{max}}}
\newcommand{\Nshell}  	{N_{\mathrm{shell}}}
\begin{document}


\title{Benchmarks of the full configuration interaction, Monte Carlo shell model, and no-core full configuration methods}

\author{T. Abe$^1$, P. Maris$^2$, T. Otsuka$^{1,3,4}$, N. Shimizu$^1$, Y. Utsuno$^5$, and J.P. Vary$^2$}

\affiliation{
$^1$ Center for Nuclear Study, the University of Tokyo, Hongo, Tokyo 113-0033, Japan \\
$^2$ Department of Physics and Astronomy, Iowa State University, Ames, Iowa 50011, USA \\
$^3$ Department of Physics, the University of Tokyo, Hongo, Tokyo 113-0033, Japan \\
$^4$ National Superconducting Cyclotron Laboratory, Michigan State University, East Lansing, Michigan 48824, USA \\
$^5$ Advanced Science Research Center, Japan Atomic Energy Agency, Tokai, Ibaraki 319-1195, Japan
}

\date{\today}

\begin{abstract}

We report no-core solutions for properties of light
nuclei with three different approaches in order to assess the accuracy
and convergence rates of each method.  Full configuration interaction
(FCI), Monte Carlo shell model (MCSM), and no core full configuration
(NCFC) approaches are solved separately for the ground state energy
and other properties of seven light nuclei using the realistic JISP16
nucleon-nucleon interaction.  The results are consistent among the
different approaches.  The methods differ significantly in how the
required computational resources scale with increasing particle number
for a given accuracy.

\end{abstract}
\maketitle

\section{Introduction and Motivation
\label{Sec_1}}

{\it Ab initio} approaches to nuclear structure and reactions for
$p$-shell nuclei have advanced significantly in the last few
years~\cite{GFMC, NCSM12,CC}.  At the same time, fundamental
approaches to the nucleon-nucleon ($NN$) and three-nucleon ($NNN$)
interactions, such as meson-exchange theory and chiral effective field
theory, have yielded major
advances~\cite{Wiringa:1994wb,Pieper_3NF,Illinois,Epelbaum,N3LO}.
Successful realistic $NN$ interactions from inverse scattering have
also emerged~\cite{Shirokov07}. These advances in microscopic nuclear
theory combine to place serious demands on available computational
resources for achieving converged properties of $p$-shell nuclei.  In
order to access a wider range of nuclei and experimental observables,
while retaining predictive power, we require
additional major advances in many-body methods.

These considerations motivate us to investigate the 
no-core Monte Carlo shell model (MCSM) which has
advantageous scaling properties for accessing larger basis spaces and
heavier nuclei.  The MCSM was first introduced in
Ref.~\cite{Otuska_MCSM} and we extend it here to treat systems without
a core. In the present work we evaluate properties of a set of
$p$-shell nuclei using the no-core MCSM and compare
with exact results in the same single-particle basis from the full
configuration interaction (FCI) method when feasible.  We also compare
with representative results from the full space {\it ab initio} no
core full configuration (NCFC)~\cite{Maris09_NCFC} method.  We adopt
the JISP16 $NN$ interaction~\cite{Shirokov07} without renormalization
and without any $NNN$ interactions.

For each of the three many-body methods, all $A$ nucleons in the
nucleus are treated on the same footing.  Experimental observables are
obtained from $A$-nucleon wave functions resulting from Hamiltonian
diagonalization in the chosen many-body basis space.  To perform the
comparisons among the methods, we focus on ground state properties of
seven nuclei as well as the properties of two low-lying narrow excited
states.

For each method, we adopt the harmonic oscillator (HO) single-particle
basis.  We obtain eigensolutions of the nuclear intrinsic Hamiltonian
expressed as a superposition of Slater determinants in the HO basis
(FCI and NCFC) or the total angular momentum projected and parity
projected deformed Slater determinants (MCSM).  Neutron and proton
orbitals are treated independently. The resulting calculated ground
state energy is a rigorous upper bound on the exact result at any
truncation.  This upper bound character applies to the lowest
calculated state of each total angular momentum and parity.

A major distinction among the methods is the definition of the cutoff
that defines the finite many-body basis space in which the
calculations are performed.  All three methods should approach the
exact solutions as the cutoffs are removed. Both the MCSM and the FCI
methods employ a cutoff in the single-particle basis $\Nshell$ which
is the highest shell of the symmetric three-dimensional HO that is
included.  All many-body basis states consistent with that cutoff are
retained (FCI) or stochastically sampled (MCSM).  On the other hand,
the NCFC approach represents an extrapolation to the infinite matrix
limit of a sequence of calculations in many-body basis spaces defined
by a many-body basis cutoff $\Nmax$, the maximum number of HO quanta
included in a many-body basis state above the minimum for that
nucleus.

A further distinction among the methods emerges from these different
truncations --- the NCFC approach may, in principle (though this is not
used in the present benchmark), guarantee the factorization of the
total wave function into an intrinsic (translationally invariant) part
times a pure $0s$ HO for the center-of-mass (c.m.) motion whereas the
MCSM and FCI approaches do not guarantee this factorization.  The
method of analysis introduced for the coupled cluster
method~\cite{CC_CM} implies that MCSM and FCI may factorize reasonably
well at an optimally chosen oscillator parameter so that observables
may be evaluated with minimal influence from spurious c.m. motion
effects. All other known symmetries of the intrinsic Hamiltonian are
retained in the many-body basis by each method.

The main motivation for the no-core MCSM approach is its superior
scaling properties with increasing nucleon number.  We estimate that,
for a fixed $\Nshell$ value and a fixed level of accuracy the MCSM
scales as $N_b^2 \times \Nsp^{3\sim4}$ where $N_b$ is
the number of Monte Carlo basis states generated in the sampling and
$\Nsp$ is the number of HO single-particle states included by
$\Nshell$.  To obtain a fixed accuracy with increasing nucleon number $A$, 
$N_b$ will have to increase as some low power of $A$, estimated at $1.5\sim2.5$ 
from the results we present below.  Assuming $N_b$ dominates the 
$A$ dependence for fixed accuracy, which seems reasonable, we 
estimate that MCSM scales as $A^{3\sim5}$.
On the other hand, the NCFC scales as $A^{12\sim14}$
for a fixed $\Nmax$ value and the maximum $\Nmax$ value roughly fixes
the accuracy of the final NCFC result.  Since the MCSM scaling for
fixed accuracy is far less dependent on the number of nucleons $A$, it
will be the superior approach once $A$ increases to the point where
the NCFC fails to generate a sufficiently converged result.
Nevertheless, the truncated calculations within NCFC will continue to
produce a valid upper bound to the exact answer.

Since the MCSM approximates the FCI calculation by stochastically
sampling the FCI many-body basis space, we provide comparisons between
these two methods in smaller basis spaces and for lighter systems
where we can still perform the FCI calculations. For these test
problems, we find that the MCSM provides an accurate approximation to
the FCI results.  The sequence of MCSM results with increasing
$\Nshell$ and for heavier nuclei may also be compared with the
sequence of results as a function of $\Nmax$ that underlie the NCFC
result in order to assess convergence rates and uncertainties in
extrapolated results.

The outline of this paper is as follows. After the Introduction and
Motivation of Sec.~\ref{Sec_1}, many-body basis space truncations and
quantum many-body methods adopted for the benchmark in this paper are
briefly described in Sec.~\ref{Sec_2}.  The selections of the $NN$
interaction and nuclear states are summarized in Sec.~\ref{Sec_3}.
The benchmark comparisons are presented and discussed in
Sec.~\ref{Sec_4}.  The summary and outlook can be found in
Sec.~\ref{Sec_5}.  In the Appendix we present additional details for
the energy variance and the extrapolation of the no-core MCSM results to
the FCI basis.

\section{Quantum Many-Body Methods Adopted
\label{Sec_2}}

A long-standing goal of nuclear physics is to obtain the exact
solutions of a realistic Hamiltonian (i.e., one that describes well the
few-body data) for finite nuclei and to compare those results with
experiment where available.  Once validated, the same methods with the
same Hamiltonian will be very useful for predicting properties of
nuclei that cannot be studied experimentally but may be of great
importance in understanding astrophysical phenomena or for practical
applications such as energy generation.  This is the physics program
we aim to empower by developing and testing new many-body methods.

We begin by introducing the elements that the three methods we study
here have in common.  The translationally invariant nonrelativistic
nuclear plus Coulomb interaction Hamiltonian is taken to consist of
\begin{equation}
 H = T_{\hbox{\scriptsize rel}} + V_{NN}  + V_{NNN} + \ldots + V_{\hbox{\scriptsize Coulomb}},
\end{equation}
where $T_{\hbox{\scriptsize rel}}$ is the internal (``relative") kinetic energy of the
nucleons and the $NN$ and $NNN$ interactions are included along with the
Coulomb interaction between the protons.  The Hamiltonian may include
additional terms such as multinucleon interactions among more than
three nucleons simultaneously and higher-order electromagnetic
interactions such as magnetic dipole-dipole terms.

The JISP16 $NN$ interaction adopted here produces a high-quality
description of the $NN$ scattering data and the
deuteron~\cite{Shirokov07} as well as a good description of a range of
properties of light nuclei~\cite{Maris09_NCFC}.  For the present
effort we neglect all other interaction terms such as the $NNN$,
higher-body strong interactions and the Coulomb interaction though the
three methods are capable of including them. These additional terms
will be required for precision descriptions of nuclear properties but
are not expected to alter the conclusions from our benchmarks here.

All calculations are performed in an $M$-scheme basis where the
many-body basis states are constructed with good total magnetic
projection $M$.  The MCSM projects out states of fixed total angular
momentum and parity $J^\pi$.  The basis states used in the FCI and
NCFC calculations are constructed with a fixed parity (as well as
fixed $M$).  The eigensolutions of the FCI and NCFC methods will
also possess good $J$ up to numerical errors.  Evaluating the value of $J$
for any eigensolution serves as a crosscheck on the precision of the
calculations.

In all applications here, we seek to obtain only the lowest few
eigenvalues and eigenfunctions.  For the NCFC and the FCI calculations
we employ the code ``Many-Fermion Dynamics - nuclear" or
``MFDn"~\cite{Vary92_MFDn} which has been optimized for
leadership-class parallel computers~\cite{Maris-ICCS}.  For the MCSM
calculations, we employ a new MCSM code that runs efficiently on
parallel computers~\cite{ref5}.

All solutions will have a dependence on the cutoff (either $\Nshell$
for FCI and MCSM or $\Nmax$ for truncated NCFC) and dependence on the
HO energy $\hbar\omega$.  The MCSM results also depend, in principle, on the
number of Monte Carlo basis states $N_b$ and we employ an extrapolation based on energy-variance 
to estimate the $N_b$-independent solution.  The degree to which we
obtain results independent of the cutoff and of the HO energy is a
measure of the convergence of the results --- fully converged results
are independent of all basis space parameters.

\subsection{Many-body basis space truncations}

The methods we investigate employ one of two different truncation
schemes as mentioned above.  The MCSM and FCI employ an $\Nshell$
cutoff while the NCFC employs $\Nmax$ to define the finite basis
spaces in which the Hamiltonian is evaluated and diagonalized.  We
work in a neutron-proton scheme rather than a basis of good isospin.
We now discuss some additional features of those truncation schemes.

\subsubsection{$\Nshell$}

For the MCSM and FCI methods, all single-particle states for neutrons
and protons in HO shells up to and including $\Nshell$ are included
($\Nshell$ = 1 for the lowest shell).  Then, all many-body states
consistent with that cutoff and the selected symmetries are
enumerated.  Thus, for example, we include basis states where all
nucleons occupy the highest HO shell if that shell can accommodate all
of them. Table~\ref{Table:Nshell} presents many-body basis space
dimensions in the $M$ scheme and $J$ scheme over a range of $\Nshell$
values for the nuclei we investigate. We also include $^{16}$O for
illustrative purposes.  An FCI calculation involves evaluating the
Hamiltonian with that dimension and diagonalizing it --- at least to
obtain the low-lying solutions of interest.

\begin{table*}[tb]
      \begin{tabular}{crrrrrr}
   \hline
   \hline
 $\Nshell$  &~~~~$2$~~~&~~~~~$3$~~~~~&~~~~~$4$~~~~~&~~~~~$5$~~~~~&~~~~~$6$~~~~~&~~~~~$7$~~~~~ \\
   \hline
 & \multicolumn{6}{c}{$M$ scheme} \\
   \hline
    $^4$He        & $\qquad$  98   & $\qquad$  3.06 $\times$ $10^3$   & $\qquad$  3.98 $\times$ $10^4$   & $\qquad$  3.14 $\times$ $10^5$   & $\qquad$  1.77 $\times$ $10^6$   & $\qquad$  7.84 $\times$ $10^6$    \\
    $^6$He        &   216  &   6.51 $\times$ $10^4$   &  3.86 $\times$ $10^6$   &   9.80 $\times$ $10^7$   &   1.45 $\times$ $10^9$   &   1.47 $\times$ $10^{10}$    \\
  $^6$Li          &   293   &   8.59 $\times$ $10^4$   &   5.08 $\times$ $10^6$   &   1.29 $\times$ $10^8$   &   1.91 $\times$ $10^9$   &   1.94 $\times$ $10^{10}$    \\
    $^7$Li          &   400   &   3.60 $\times$ $10^5$   &   4.51 $\times$ $10^7$   &   2.05 $\times$ $10^9$   &   4.91 $\times$ $10^{10}$   &   7.50 $\times$ $10^{11}$    \\
    $^8$Be         &   518   &   1.47 $\times$ $10^6$   &   3.96 $\times$ $10^7$   &   3.24 $\times$ $10^{10}$   &   1.26 $\times$ $10^{12}$   &   2.91 $\times$ $10^{13}$     \\
    $^{10}$B      &   293   &   1.34 $\times$ $10^7$   &   1.82 $\times$ $10^{10}$   &   5.02 $\times$ $10^{11}$   &   5.22 $\times$ $10^{14}$   &   2.78 $\times$ $10^{16}$    \\
    $^{12}$C     &   98   &   8.22 $\times$ $10^7$   &   5.87 $\times$ $10^{11}$   &   5.50 $\times$ $10^{14}$   &   1.54 $\times$ $10^{17}$   &   1.90 $\times$ $10^{19}$   \\
    $^{16}$O     &    1    & 8.12 $\times$ $10^8$  &  2.10 $\times$ $10^{14}$  &  2.51 $\times$ $10^{18}$  &  5.32 $\times$ $10^{21}$ &  3.59 $\times$ $10^{24}$ \\
   \hline
 & \multicolumn{6}{c}{$J$ scheme} \\
   \hline
  $^4$He        &   20   &   2.72 $\times$ $10^2$ &   2.10 $\times$ $10^3$   &   1.12 $\times$ $10^4$  &   4.58 $\times$ $10^4$   &   1.54 $\times$ $10^5$    \\
  $^6$He        &   35   &   3.93 $\times$ $10^3$   &   1.37 $\times$ $10^5$   &   2.35 $\times$ $10^6$  &   2.52 $\times$ $10^7$   &   1.93 $\times$ $10^8$    \\
    $^6$Li          &   97   &   1.42 $\times$ $10^4$   &   5.19 $\times$ $10^5$   &   9.05 $\times$ $10^6$  &   9.79 $\times$ $10^7$   &   7.57 $\times$ $10^8$   \\
    $^7$Li          &   89   &   3.63 $\times$ $10^4$   &   2.73 $\times$ $10^6$   &   8.40 $\times$ $10^7$  &   1.46 $\times$ $10^9$   &   1.69 $\times$ $10^{10}$    \\   
    $^8$Be        &   70   &   6.89 $\times$ $10^4$   &   1.08 $\times$ $10^7$   &   5.92 $\times$ $10^8$  &   1.66 $\times$ $10^{10}$   &   2.90 $\times$ $10^{11}$    \\ 
    $^{10}$B     &   43   &   2.20 $\times$ $10^6$   &   1.21 $\times$ $10^9$   &   2.21 $\times$ $10^{11}$  &   1.65 $\times$ $10^{13}$   &   6.67 $\times$ $10^{14}$    \\
    $^{12}$C     &   20   &   2.94 $\times$ $10^6$   &   1.14 $\times$ $10^{10}$   &   6.94 $\times$ $10^{12}$  &   1.38 $\times$ $10^{15}$   &   1.28 $\times$ $10^{17}$    \\
    $^{16}$O     &   1   &   2.54 $\times$ $10^7$   &   3.26 $\times$ $10^{12}$   &   2.46 $\times$ $10^{16}$  &   3.66 $\times$ $10^{19}$   &   1.84 $\times$ $10^{22}$    \\
   \hline
   \hline
  \end{tabular}
\caption{Dimensions of the $M$ scheme (top) and $J$ scheme (bottom)
  many-body basis spaces for selected nuclei with the $\Nshell$
  truncation. The dimensions are for the natural parity states with
  $M$ and $J$ taken to be the lowest allowed value ($M = 0$ for even
  nuclei except for $^6$Li and $^{10}$B where $M = 1$; $M = 1/2$ for
  odd nuclei, and similarly for $J$). 
  \label{Table:Nshell}}
\end{table*}

\subsubsection{$\Nmax$}

For the NCFC method, we employ the many-body $\Nmax$ truncation where
we enumerate all many-body states, with the selected symmetries,
possessing total HO quanta less than or equal $\Nmax$ above the lowest
allowed configuration for that nucleus.  Each single-particle state in a basis state
contributes $2n+l$ to the total HO quanta ($n$ is the radial quantum
number and $l$ is the orbital angular momentum quantum number) for
that basis state and then the minimum sum for that nucleus is
subtracted to give the total quanta above the minimum for that basis
state. The basis space for each nucleus begins with $\Nmax=0$ and
increases in units of $2$ for the natural parity states.  Odd values
of $\Nmax$ cover the unnatural parity states.  Thus, for example, we
include basis states where one nucleon occupies the highest HO shell
accessed. Table~\ref{Table:Nmax} presents many-body basis space
dimensions in the $M$ scheme over a range of $\Nmax$ values for the
nuclei we investigate, again with $^{16}$O added for illustrative
purposes.  A no-core shell model (NCSM) calculation involves
evaluating the Hamiltonian with that dimension and diagonalizing it -
at least to obtain the low-lying solutions of interest.  A sequence 
with increasing $\Nmax$ of NCSM calculations using a Hamiltonian 
defined for the infinite basis will converge from above to the exact solution. 
The NCFC approach uses that sequence to extrapolate to the infinite basis limit.

\begin{table*}[tb]
      \begin{tabular}{crrrrrrrr}
   \hline 
   \hline 
\multicolumn{1}{c}{ $\Nmax$}  & \multicolumn{1}{c}{ $0$ } & \multicolumn{1}{c}{ $2$ } & \multicolumn{1}{c}{ $4$ } & \multicolumn{1}{c}{ $6$ } & \multicolumn{1}{c}{ $8$ } & \multicolumn{1}{c}{ $10$ } & \multicolumn{1}{c}{ $12$ } & \multicolumn{1}{c}{ $14$ } \\
   \hline
    $^4$He        & $\quad$  1   & $\quad$  5.9 $\times$ $10^1$   & $\quad$  9.52 $\times$ $10^2$  & $\quad$  7.92 $\times$ $10^3$   & $\quad$  4.48 $\times$ $10^4$   & $\quad$  1.96 $\times$ $10^5$   & $\quad$  7.14 $\times$ $10^5$  & $\quad$ 2.25 $\times$ $10^6$  \\
    $^6$He        &   5   &   5.11 $\times$ $10^2$  &   1.17 $\times$ $10^4$   &   1.40 $\times$ $10^5$   &   1.14 $\times$ $10^6$   &   7.06 $\times$ $10^6$   &   3.58 $\times$ $10^7$   &  1.56 $\times$ $10^8$   \\
    $^6$Li          &    8   &   7.11 $\times$ $10^2$  &   1.58 $\times$ $10^4$   &   1.87 $\times$ $10^5$   &   1.51 $\times$ $10^6$   &   9.36 $\times$ $10^6$   &   4.75 $\times$ $10^7$   &  2.06 $\times$ $10^8$   \\
    $^7$Li          &   21   &   1.96 $\times$ $10^3$  &   4.89 $\times$ $10^4$   &   6.64 $\times$ $10^5$   &   6.15 $\times$ $10^6$   &   4.36 $\times$ $10^7$   &   2.52 $\times$ $10^8$   &  1.24 $\times$ $10^9$   \\
    $^8$Be          &   51   &   5.10 $\times$ $10^3$   &   1.44 $\times$ $10^5$   &   2.22 $\times$ $10^6$   &   2.33 $\times$ $10^7$   &   1.87 $\times$ $10^8$   &   1.22 $\times$ $10^9$   &  6.77 $\times$ $10^9$   \\
    $^{10}$B      &   73   &   1.35 $\times$ $10^4$   &   5.51 $\times$ $10^5$   &   1.16 $\times$ $10^7$   &   1.60 $\times$ $10^8$   &   1.65 $\times$ $10^9$   &   1.40 $\times$ $10^{10}$   &  9.63 $\times$ $10^{10}$  \\
    $^{12}$C     &   51   &   1.77 $\times$ $10^4$   &   1.12 $\times$ $10^6$   &   3.26 $\times$ $10^7$   &   5.94 $\times$ $10^8$   &   7.83 $\times$ $10^9$   &   8.08 $\times$ $10^{10}$   &  6.88 $\times$ $10^{11}$   \\
    $^{16}$O     &   1   &   1.25 $\times$ $10^3$   &   3.45 $\times$ $10^5$   &   2.65 $\times$ $10^7$   &   9.97 $\times$ $10^8$   &   2.37 $\times$ $10^{10}$   &   4.06 $\times$ $10^{11}$   &  5.43 $\times$ $10^{12}$   \\
   \hline
   \hline
  \end{tabular}
\caption{Dimensions of the $M$-scheme many-body basis spaces for
  selected nuclei with $\Nmax$ truncation. The dimensions are for
  the natural parity states with $M$ taken to be the lowest allowed
  value ($M = 0$ for even nuclei except for $^6$Li and $^{10}$B where
  $M = 1$; $M = 1/2$ for odd nuclei).  The
  sequence of dimensions for unnatural parity states (odd values of
  $\Nmax$) lie intermediate the neighboring natural parity
  dimensions.
  \label{Table:Nmax}}
\end{table*}

\subsection{FCI}

An FCI calculation involves solving the Hamiltonian eigenvalue problem
in a many-body basis space with the $\Nshell$ truncation described
above.  We have performed sets of these calculations in the present
effort to provide the exact results for comparison with the MCSM
approach and to compare with the $\Nmax$-truncated results of the NCFC
approach.  For the FCI results reported here, we employ the $M$-scheme
basis whose dimensions are indicated in Table~\ref{Table:Nshell} and
use the Lanczos algorithm in a manner similar to a NCSM calculation.
Unlike the NCFC approach, we do not perform an extrapolation to the
infinite matrix limit of the FCI results as a function of $\Nshell$.

\subsection{MCSM}

The MCSM approach~\cite{Otuska_MCSM,ref5} proceeds through a sequence
of diagonalization steps within the Hilbert subspace spanned by the
selected importance-truncated bases, beginning with, in principle, any
initial trial solution for the system. Until now, the deformed
Hartree-Fock (Hartree-Fock-Bogoliubov) states in the HO
single-particle basis defined by the $\Nshell$ cutoff have been adopted as
an initial state for the shell-model calculations with a core in light
(medium-heavy) nuclei.  These deformed single-particle states in a
canonical basis are constructed as a linear combination of spherical
HO single-particle states up to and including those in the $\Nshell$
cutoff. One then stochastically samples all possible many-body basis
states around the mean field solutions with the aid of the auxiliary
fields and diagonalizes the Hamiltonian matrix within the subspace
spanned by these bases. An accept/reject process of a stochastically
sampled basis is performed by minimizing the energy variationally, not
by the importance sampling in quantum Monte Carlo methods implemented
by the Metropolis algorithm. The MCSM is thus not the usual ``quantum"
Monte Carlo, but can evade the so-called negative sign problem, which
is the fundamental issue that cannot be avoided in quantum Monte Carlo
methods. 

In the MCSM, a many-body state $|\Psi^{J^\pi M} \rangle$ is
constructed from the linear combination of non-orthogonal
angular-momentum ($J$) and parity ($\pi$) projected deformed Slater
determinants $| \Phi \rangle$ with good total angular momentum
projection ($M$) as a stochastically selected basis,
\begin{equation}
|\Psi^{J^\pi M} \rangle = \sum_{n=1}^{N_b} f_n | \Phi_n^{J^\pi M} \rangle, 
\end{equation} 
where the angular-momentum and parity projected basis, 
\begin{equation}
| \Phi^{J^\pi M} \rangle = \sum_{K=-J}^{J} g_K P_{MK}^J P^\pi | \phi \rangle, 
\end{equation}
and the deformed Slater determinant,
\begin{equation}
| \phi \rangle = \prod_{i=1}^{A} a_i^\dagger | - \rangle, 
\end{equation}
with the vacuum $| - \rangle$ and the creation operator $a_i^\dagger =
\sum_{\alpha=1}^{\Nsp} c_{\alpha}^\dagger D_{\alpha i}$.  The
coefficient $D_{\alpha i}$ is stochastically sampled by the
auxiliary-field Monte Carlo technique around the Hartree-Fock
solutions.

With increasing Monte Carlo basis dimension, the ground state energy
of a MCSM calculation converges from above to the exact value --- the
value that would be obtained by diagonalization of the corresponding
FCI basis space.  The energy, therefore, always gives the variational
upper bound in this framework.

An exploratory no-core MCSM investigation of the proof-of-the
principle type has been done for the low-lying states of the Berylium
isotopes by applying the existing MCSM algorithm with a core to a
no-core problem~\cite{Liu2011}.  Recent improvements
on the MCSM algorithm have enabled significantly larger
calculations~\cite{ref5,Utsuno:2012vm}.  In addition energy variance
extrapolation methods have been introduced and tested in order to
obtain precise results at each $\Nshell$ cutoff~\cite{ref6}.  We adopt
these improvements in the present work, and extend our earlier
investigations~\cite{ref7}.  A similar work by the hybrid
multideterminant method is also proposed \cite{Puddu}.

\subsection{NCFC}

The NCFC approach~\cite{Maris09_NCFC} aims to achieve the solution of
the nuclear many-body problem by diagonalization in a sufficiently
large basis space that converged energies are accessed ---
either directly or by simple extrapolation.  Convergence is assessed
in the two-dimensional parameter space of the basis space
($\hbar\omega$,~$\Nmax$) and is defined as independence of both
parameters within estimated uncertainties.  Each observable is studied
independently to obtain its converged value and its assessed
uncertainty.

The NCFC is both related to and distinct from the NCSM~\cite{NCSM12}
that features a finite matrix truncation and an effective Hamiltonian
renormalized to that finite space.  The $\Nmax$ regulator appears in
both the NCFC, where it is taken to infinity, and in the NCSM, where
it also appears in the definition of the effective Hamiltonian.  In
both the NCFC and the NCSM, the $\Nmax$ cutoff in the HO basis is
needed to preserve Galilean invariance --- to factorize all solutions
into a product of internal motion and c.m. motion components.  
With $\Nmax$ as the regulator, both the NCFC and the NCSM are
distinguished from the FCI and the MCSM approaches where such
factorization is not guaranteed but may be approximately
valid~\cite{CC_CM}.  In the NCFC approach the ground state energies at
any finite truncation are strict upper bounds, and converge
monotonically to the exact result.  This facilitates a simple
extrapolation.  The approach to the exact result is in general not
monotonic in the NCSM. 

\subsection{Snapshot comparison}

\begin{figure}[tbp]
\begin{center}
\includegraphics[width=102mm]{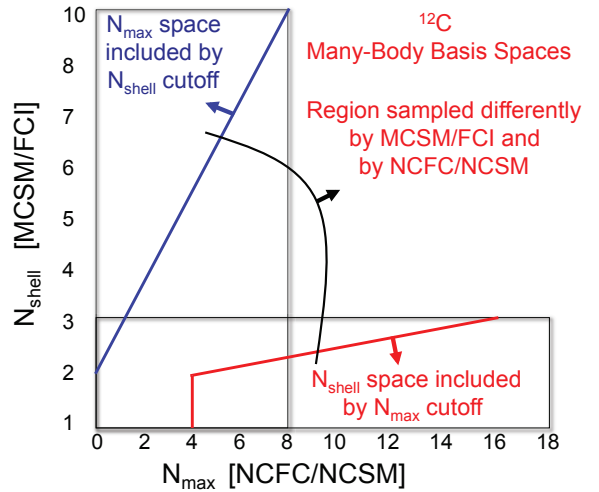}
\caption{(Color online) Overview of the basis spaces covered with the
  many-body methods discussed here for the case of $^{12}$C.
  $\Nmax$ is defined as the number of oscillator quanta above lowest
  possible number of quanta.  $\Nshell$ is the number of oscillator
  shells counting the $0s$ shell as the first shell.  The MCSM
  incorporates an FCI space.  That is, all single-particle states in
  the included shells are available to all particles without
  additional restrictions except for symmetry constraints.}
\label{Fig:Basis_spaces}
\end{center}
\end{figure}

For convenience, we present simple comparisons among the methods we
employ in Fig.~\ref{Fig:Basis_spaces} and in
Table~\ref{snap_shot_comparison}.
In Fig.~\ref{Fig:Basis_spaces} we show, for the specific example of
$^{12}$C, how the many-body basis spaces both overlap and differ from
each other as a function of increasing cutoff.  To indicate an area of
complete overlap, the red curve in Fig. \ref{Fig:Basis_spaces} borders
the $\Nshell$ space included as a function of increasing $\Nmax$. On
the other hand, the blue curve borders the region of $\Nmax$ space
included as a function of increasing $\Nshell$.

\begin{table}[tb]
   \begin{tabular}{lccc}
\hline 
\hline 
   snapshot comparison  & $\quad$ FCI $\quad$           &  $\quad$ MCSM $\quad$        & $\quad$ NCFC $\quad$     \\
   \hline
   c.m. motion             & approx.        & approx.      & exact     \\
   Spectra               & OK             & some         & OK        \\
   wfns  $\rightarrow$ observables & \checkmark     & \checkmark   & \checkmark\\
   Matrix dimension      & $\alt 10^{10}$  & $\alt 10^{20}$ & $\alt 10^{10}$ \\
   Scaling with $A$      &  $A^{18\sim20}$   & $A^{3\sim5}$  &  $A^{12\sim14}$ \\
   No. parallel cores    &   $10^5$       &   $10^5$      & $10^5$   \\
   Comp'l bottleneck     & Memory         & CPU time      & Memory \\
\hline 
\hline 
  \end{tabular}
  \caption{
    Overview of the current features of the three 
    no-core many-body methods employed in this work.  The estimates of
    the scaling with the number of nucleons $A$ are very crude and
    based on applications to light ($p$-shell) nuclei for a fixed accuracy. The last two
    lines in the table present overall characteristic features of the
    codes used for this work.
    \label{snap_shot_comparison}}
\end{table}

Table~\ref{snap_shot_comparison} presents a simple set of comparisons
and contrasts between the methods. Since $\Nshell$ and $\Nmax$ 
roughly signify respective accuracies of the methods we hold them fixed 
to facilitate these comparisons. We emphasize that these are the
current features and limitations of these approaches.  Additional
developments underway are aimed at improving each method, especially
the MCSM and NCFC approaches.  

The main advantage of the MCSM approach is that, at fixed $\Nshell$,
the increase in computational needs with increasing nucleon number
(``scaling" in Table \ref{snap_shot_comparison}) of the MCSM approach
is much slower than that of the FCI appoach.  In addition, the increase in
computational needs with $A$ of the MCSM approach at fixed $\Nshell$
is significantly slower than that of NCFC at fixed $\Nmax$.  
Note that the MCSM algorithm is CPU bound, and may be suitable for
implementation on general purpose graphical processing units (GPGPUs).

Before progressing to the detailed comparisons among the results from
the methods we investigate here, it is worth noting that there are
additional efforts aimed at accelerating the convergence of 
{\it ab initio} no-core many-body methods using basis function
techniques.  The ``Importance-Truncated" no-core shell model
(IT-NCSM)~\cite{Roth} attempts to estimate the contributions of the
many-body configurations above the $\Nmax$ cutoff using sequences of
perturbative contributions to the energy of low-lying states.  The
symmetry-adapted no-core shell model (SA-NCSM)~\cite{Dytrych} aims to
augment the basis space above the $\Nmax$ cutoff by adding basis
states of selected symmetry character that are preferred by low-lying
nuclear collective motion.  Both methods are producing impressive
results.  It remains to be seen which method, among the many under
investigation, will be more efficient and for which systems and which
observables.  Outstanding challenges include the fully microscopic
description of clustering phenomena and extensions to {\it ab initio}
nuclear reaction theory.

\section{Selections of Ingredients
\label{Sec_3}}

We have outlined above the many-body methods selected for these
benchmark comparisons (FCI, MCSM, and NCFC).  All results are obtained in a
HO basis of single-particle states characterized by the oscillator
energy $\hbar\omega$ in MeV and the cutoff of the basis space
($\Nshell$ or $\Nmax$) defined above.  We adopt the JISP16 $NN$
interaction~\cite{Shirokov07} without renormalizing it to a lower
momentum scale and we neglect Coulomb and all other interactions.  The
contributions of spurious c.m. excitation are not
discussed here in any detail.  Such contributions are absent in
conventional NCFC results for ground state observables where we would
include the standard Lagrange multiplier term that constrains the c.m.
motion to the $0s$ HO state. However, for the present benchmark
comparisons, we have dropped the Lagrange multiplier term in the
Hamiltonian for simplicity.  We do not expect that spurious c.m.
effects play a significant role in our benchmark comparisons. 

\subsection{Interaction}

The JISP16 $NN$ interaction is determined by inverse scattering
techniques from the $np$ phase shifts and is taken to be charge
independent.  JISP16 is available in a relative HO
basis~\cite{Shirokov07} and can be written as a sum over partial
waves
\begin{equation}
 V_{NN} = \sum_{S, {\cal J}, T} {\cal P}_{S, {\cal J}, T} 
 \sum_{n, \ell , n^\prime, \ell^\prime} 
 \;| n , \ell \rangle 
 \
 A^{S, {\cal J}, T}_{n\ell,n^\prime \ell^\prime}
 \
 \langle n^\prime \ell^\prime |, 
\end{equation}
where $\hbar \omega = 40$ MeV and ${\vec {\cal J}} = {\vec \ell} +
{\vec s}$.  The HO basis state of relative motion is signified by $| n
\ell \rangle$ and the projector onto the specified channel is
represented by ${\cal P}_{S, {\cal J}, T}$.  A small number of
coefficients $\{ A^{S, {\cal J}, T}_{n\ell,n^\prime \ell^\prime} \}$
are sufficient to describe the phase shifts in each partial wave.
Note that the JISP16 interaction is nonlocal and its off-shell
properties have been tuned by phase-shift equivalent transformations
to produce good properties of light nuclei.  For example, JISP16 is
tuned in the $^3S_1-^3D_1$ channel to give a high precision
description of the deuteron's properties.  Other channels are tuned to
provide good descriptions of $^3$H binding, the low-lying spectra of
$^6$Li and the binding energy of $^{16}$O. With these off-shell
tunings to nuclei with $A \geq 3$ one may view JISP16 as simulating,
to some approximation, what would appear as $NNN$ interaction
contributions (as well as higher-body interactions) in alternative
formulations of the nuclear Hamiltonian.

\subsection{Nuclear states evaluated}

For this benchmark process, we select nine states of light nuclei that
includes seven ground states and two excited states; $^4$He ($0^+$), $^6$He
($0^+$), $^6$Li ($1^+$), $^7$Li ($1/2^-$, $3/2^-$), $^8$Be ($0^+$),
$^{10}$B ($1^+$, $3^+$), and $^{12}$C ($0^+$).  We compare results for
the energy, the point-particle root-mean-square (rms) matter
radius, and the electric quadrupole and magnetic dipole moments.

Our goal here is to compare the methods at fixed finite cutoffs.  To
achieve convergence of the quantities we evaluate will require a much
larger effort than the present undertaking.  For the benchmark
process, we simply proceed through a sequence of cutoffs for each
state and each method and obtain results as a function of the
oscillator energy, $\hbar\omega$.  Then, since all our methods retain
the variational principle, we select the optimal $\hbar\omega$ that
minimizes the energy for that state and basis space cutoff.
We compare the observables for that optimal $\hbar\omega$.

The MCSM results are compared with those of FCI which gives the exact
results in the chosen single-particle model space. The FCI results are
obtained by the MFDn code~\cite{Vary92_MFDn,Maris-ICCS} and the MCSM
results by the newly developed code~\cite{ref5,Utsuno:2012vm}.  Note
that the FCI results are not available for all the cases presented
here due to computational limitations of the FCI approach as indicated
by the ``Matrix dimension" entry in Table~\ref{snap_shot_comparison}.

\section{Benchmark Comparisons
\label{Sec_4}}

\subsection{Results for energies}

\begin{figure*}[tbp]
\includegraphics[width=1.8\columnwidth]{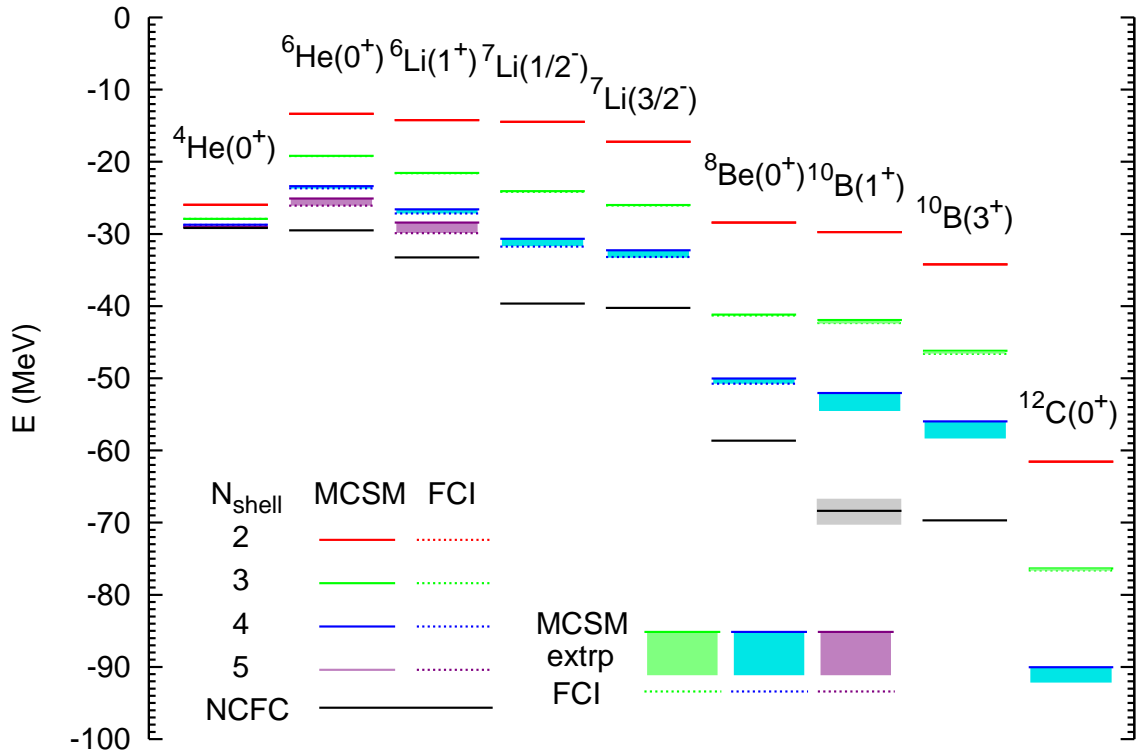}
\caption{(Color online) 
  Comparisons of the energies between the MCSM and FCI along
  with the fully converged NCFC results where available. The NCFC
  result for the $^{10}$B($1^+$) state has a large uncertainty indicated
  by the grey band. The MCSM
  (FCI) results are shown as the solid (dashed) lines that nearly
  coincide where both are available.  The extrapolated MCSM results
  are illustrated by bands.  From top to bottom, the truncation of the
  model space is $\Nshell = 2$ (red), $3$ (green), $4$ (blue), and
  $5$ (purple).  Note that the MCSM results are extrapolated by the
  energy variance with the second-order polynomials~\cite{ref6}.  
  Also note that all of the results of $^{10}$B and $^{12}$C at 
  $\Nshell=4$ were obtained only with MCSM.
  \label{fig2} }
\end{figure*}

We present the energies obtained by MCSM and FCI in
Fig.~\ref{fig2} through a sequence of $\Nshell$ truncations.  For the
$A = 4$ and $6$ systems we obtain results with both MCSM and FCI
through $\Nshell = 5$.  For the systems with $A=7$ and $A=8$ we
obtain results with both MCSM and FCI through $\Nshell=4$.  
Finally, for $^{10}$B ($1^+$, $3^+$) and $^{12}$C ($0^+$) we are not
able to obtain the FCI results beyond $\Nshell = 3$ due to
computational limitations so the MCSM results at $\Nshell = 4$
represent predictions.  That is, the FCI $M$-scheme matrix dimensions
are 18.2 billion and 587 billion as shown in Table~\ref{Table:Nshell}
for $^{10}$B and $^{12}$C, respectively, at $\Nshell = 4$ and these
dimensions exceed our current FCI capabilities.

The MCSM and FCI nearly coincide in all cases where both are
available.  In fact, for $\Nshell=2$ they are identical to within
machine precision. 
This can easily be understood because the
dimension of the complete FCI basis in the $J$ scheme is below 100
(except for the $J^\pi=\frac{3}{2}^-$ basis space of $^7$Li,
see Table~\ref{Table:Nshell}),
the number of Monte Carlo states used
in most of the calculations presented here.\footnote{The Monte Carlo
  basis states do not form an orthogonal basis, and can be
  overcomplete.}  Indeed, the MCSM results tend to become 
independent of the number of Monte Carlo states $N_b$ once $N_b$
is of the order of 10\% to 20\% of the dimension $D$ of the
underlying FCI basis in the $J$ scheme.  Also for $\Nshell=3$ the MCSM and
FCI results are virtually indistinguishable in Fig.~\ref{fig2}.  After
the extrapolation of the MCSM results using the energy
variance~\cite{ref6}, as discussed below and indicated by shaded
regions in Fig.~\ref{fig2}, we obtain also very good agreement with
the FCI results (where available) for $\Nshell \ge 4$.

In Fig.~\ref{fig2} we also present the NCFC results for the energies
of the $A=4$ through $10$ nuclei (black solid lines) for comparison.
The NCFC results are obtained from calculations up through $\Nmax=14$
for $A=4$ and $6$, and up through $\Nmax = 12$ for $A=7$ and $8$,
using an exponential extrapolation to the infinite basis space.  
In these cases, the extrapolation uncertainties in the fully converged 
NCFC results are less than the width of the black line.  For $A=10$ we
employ results up through $\Nmax = 10$ to obtain the NCFC results.
The extrapolation of the $1^+$ state in $^{10}$B to obtain the quoted 
NCFC result has a significantly greater uncertainty due to the occurrence 
of two close-lying $1^+$ states in the calculated spectrum

\begin{table*}[tbp]
    \begin{tabular}{lllcclcclcclccl}
      \hline
      \hline
 & & \multicolumn{12}{c}{ $E$ (MeV) } \\
\cline{3-15}
Nuclei & \multicolumn{1}{c}{Method} & \quad $\Nshell=2$ \quad & $\hbar\omega$ & $D$ 
                & \quad $\Nshell=3$ \quad & $\hbar\omega$ & $N_b$ 
                & \quad $\Nshell=4$ \quad & $\hbar\omega$ & $N_b$ 
                & \quad $\Nshell=5$ \quad & $\hbar\omega$ & $N_b$ & \multicolumn{1}{c}{\quad NCFC \quad } \\
      \hline
        $^4$He ($0^+$)   &MCSM  & \quad -25.956 & 30 &  20 & \quad -27.914 & 30 & 100 & \quad -28.737 & 30 & 100 & \quad -29.011       & 25 & 50 & \quad -29.164(2) \\
                         &extrp&        &    &     &         &    &     & \quad -28.738(1) &    &     & \quad -29.037(1) &    &    & \\
                         &FCI & \quad -25.956 &    &     & \quad -27.914 &    &     & \quad -28.738    &    &     & \quad -29.036    &    &    & \\
        $^6$He ($0^+$)   &MCSM& \quad -13.343 & 20 &  35 & \quad -19.186 & 20 & 100 & \quad -23.480 & 25 & 100 & \quad -25.080 & 25 & 50 & \quad -29.51(5) \\
                         &extrp&        &    &     & \quad -19.196(1) &    &     & \quad -23.687(4) &    &     & \quad -26.086(76)     &    &  &  \\
                         &FCI & \quad -13.343 &    &     & \quad -19.196    &    &     & \quad -23.684    &    &     & \quad -26.079     &    &  &  \\
        $^6$Li ($1^+$)   &MCSM& \quad -14.218 & 20 &  97 & \quad -21.549 & 20 & 100 & \quad -26.757 & 25 & 100 & \quad -28.410 & 25 & 50 & \quad -33.22(4) \\

                         &extrp&        &    &     & \quad -21.581(1) &    &     & \quad -27.166(16) &    &     & \quad -29.873(83) &    &    \\
                         &FCI & \quad -14.218 &    &     & \quad -21.581    &    &     & \quad -27.168     &    &     & \quad -29.893      &    &    \\
        $^7$Li ($1/2^-$) &MCSM& \quad -14.459 & 20 & 89  & \quad -24.073 & 20 & 100 & \quad -30.904 & 25 & 100 & & & & \quad -39.8(1) \\
                         &extrp&        &    &     & \quad -24.167(2) &    &     & \quad -31.780(51) &    &     & & & \\

                         &FCI & \quad -14.458 &    &     & \quad -24.165    &    &     & \quad -31.748     &    &     & & & \\
        $^7$Li ($3/2^-$) &MCSM& \quad -17.232 & 20 & 130 & \quad -25.978 & 25 & 100 & \quad -32.494 & 25 & 100 & & & & \quad -40.4(1) \\
                         &extrp&        &    &     & \quad -26.061(4) &    &     & \quad -33.272(89) &    &     & & & \\
                         &FCI & \quad -17.232 &    &     & \quad -26.063     &    &     & \quad -33.202     &    &     & & & \\
        $^8$Be ($0^+$)   &MCSM& \quad -28.435 & 20 &  70 & \quad -41.242 & 25 & 100 & \quad -50.222 & 25 & 100 & & &  & \quad -59.1(1) \\
                         &extrp&        &    &     & \quad -41.293(1) &    &     & \quad -50.753(32) &    &     & & & \\
                         &FCI & \quad -28.435 &    &     & \quad -41.291    &    &     & \quad -50.756     &    &     & & & \\
      $^{10}$B ($1^+$)   &MCSM& \quad -29.755 & 25 &  43 & \quad -41.965 & 25 & 100 & \quad -52.239 & 25 & 100 & & & & \quad -68.5(1.5)\\
                         &extrp&        &    &     & \quad -42.357(46) &    &     & \quad -54.89(16) &    &     & & & \\
                         &FCI & \quad -29.755 &    &     & \quad -42.338     &    &     &              &    &     & & & \\
      $^{10}$B ($3^+$)   &MCSM& \quad -34.221 & 25 & 97 & \quad -46.263 & 25 & 100 & \quad -56.346 & 25 & 100 & & & & \quad -69.8(2) \\
                         &extrp&        &    &     & \quad -46.618(22) &    &     & \quad -58.41(13) &    &     & & & \\
                        &FCI & \quad -34.221 &    &     & \quad -46.602     &    &     &              &    &     & & & \\
      $^{12}$C ($0^+$)   &MCSM& \quad -62.329 & 30  & 20 & \quad -76.413 & 30 & 100 & \quad -90.158 & 30 & 100 & & & \\
                         &extrp&        &    &     & \quad -76.621(4) &    &     & \quad -91.957(43) &    &     & & & \\
                         &FCI & \quad -62.329 &    &     & \quad -76.621    &    &     &             &    &     & & & \\
      \hline
      \hline
    \end{tabular}
\caption{Energies in MeV calculated for the seven ground states
  and two excited states within the MCSM and FCI methods using the
  JISP16 $NN$ interaction.  The entries of the MCSM indicate the MCSM
  results before the energy variance extrapolation, while the those of
  the ``extrp'' line denote the MCSM results after the extrapolations.
  The number of Monte Carlo vectors evaluated in the MCSM approach is
  indicated by $N_b$.  The value cited for $\hbar\omega$ (units are
  MeV) represents the value at which the energy for that state
  reaches its minimum value in that $\Nshell$ basis.  Uncertainties
  in extrapolated results are quoted in parenthesis. 
  \label{Binding_energies}}
\end{table*}
%
In order to stimulate future comparisons with other many body methods,
we present detailed results in tables for selected values of
$\Nshell$.  For the energies we present results according to the
method and the basis space cutoff in Table~\ref{Binding_energies}.
All results are presented for the value of $\hbar\omega$ where that
state is a minimum in that $\Nshell$ basis, except for the NCFC
results, which are, within the estimated numerical uncertainty,
independent of any basis parameters.  Here, we observe that the
differences between the MCSM and FCI results is at most a few hundred
keV for $\Nshell=3$, which is why they are barely distinguishable at
the energy scale of Fig.~\ref{fig2}.  For $\Nshell > 3$, this
difference can be of the order of an MeV or more.  However,
extrapolated MCSM results agree with the FCI results to within the
estimated extrapolation error, with only one case in which the
difference is larger than the estimated extrapolation error in
Table~\ref{Binding_energies}.  That case is $^{8}$Be at $\Nshell =4$
where the uncertainty is $1$ keV and the difference is $2$ keV.

The energies converge uniformly from above as expected with increasing
$\Nshell$.  We obtain significant increases in binding with each
increment in $\Nshell$ and this encourages us to develop the MCSM
further in order to access larger $\Nshell$ bases.  At the present
time, our limited results do not indicate a pattern that we can
extrapolate to the infinite $\Nshell$ limit.  However, the expected
outcomes of such extrapolations should be the NCFC results shown in
Fig.~\ref{fig2} and in the last column of
Table~\ref{Binding_energies}.  Larger $\Nshell$ results and
extrapolations to the infinite $\Nshell$ limit constitute goals for
future efforts since our main goal here is to benchmark the MCSM
approach through the range of $\Nshell$ values accessible by FCI and
to compare with the fully converged NCFC where available.

\begin{figure*}[tbp]
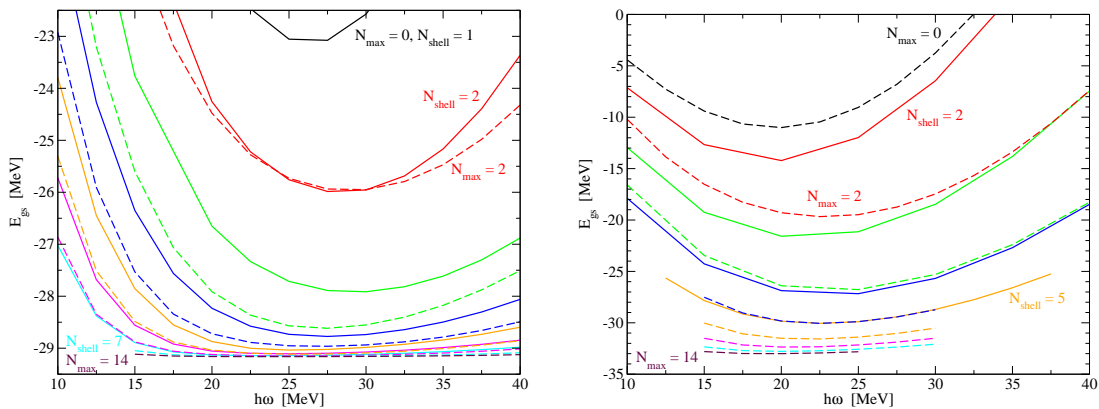

\includegraphics[width=0.80\columnwidth]{results_4He.eps}\qquad\includegraphics[width=0.80\columnwidth]{results_6Li.eps}
\caption{(Color online) 
  Comparisons of the ground state energies of $^4$He (left) and $^6$Li
  (right) between the FCI with $\Nshell$ truncation and NCFC with
  $\Nmax$ truncation as a function of the HO frequecy
  $\hbar\omega$. The FCI (NCFC with cutoff) results are shown as the
  solid (dashed) lines.  From top to bottom, the truncation of the
  model space increments by unity for $\Nshell$ up to $\Nshell=7$
  (cyan) for $^4$He up to $\Nshell=5$ (orange) for $^6$Li, and by two
  for $\Nmax$ up to $\Nmax=14$ (purple) for both $^4$He and $^6$Li.
  \label{FCI_hw}}
\end{figure*}

The detailed convergence pattern for ground state energy for $^4$He is
shown for the FCI and NCFC methods in the left panel of
Fig.~\ref{FCI_hw} as a function of $\hbar\omega$ and the basis space
cutoff ($\Nshell$ for FCI; $\Nmax$ for the approach to NCFC).  We
define convergence as independence of both $\hbar\omega$ and the basis
space cutoff.  We note that FCI at $\Nshell=8$ and the NCFC truncated
at $\Nmax=10$ both yield almost the same ground state energy of $-29.15$
MeV, even though the dimensions are quite different: the full
$\Nshell=8$ basis space dimension of $^4$He is $29 \ 031 \ 044$, whereas the
$\Nmax=10$ basis space dimension is only $196 \ 438$, more than two orders
of magnitude smaller.  The NCFC result for $^4$He is $-29.164(2)$ MeV. For comparison the MCSM results at 
$\Nshell=5$ (the largest MCSM space reported here) and $\hbar \omega=25$ MeV, 
once extrapolated with the energy-variance method, produces $-29.037(1)$ MeV 
which agrees to within 1 keV of the FCI result for that space

We also show the detailed convergence patterns for the ground state
energy of $^6$Li in the right panel of Fig.~\ref{FCI_hw} for FCI and
NCFC at various truncations as a function of $\hbar\omega$ and the
basis space cutoff.  We note that the convergence trends of the
$\Nshell = 2 \rightarrow 5$ results for $^6$Li shown in the right
panel of Fig.~\ref{FCI_hw} has both similarities and differences from
the pattern for $^4$He seen in the left panel.  Both exhibit the ``U"-shaped patterns for each truncation with the bottom of the ``U"
becoming flatter as the cutoff increases.  However, for $^4$He, the
minimum with respect to $\hbar\omega$ for $\Nshell \geq 1$ remains at
nearly a constant $\hbar\omega$ value as the cutoff is removed while
for $^6$Li that minimum shifts to higher values of $\hbar\omega$.

For $^6$Li, the ground state energy increments by about the same amount
from $\Nshell = 3 \rightarrow 4$ as for $\Nshell = 2 \rightarrow 3$.
However, there is a significant decrease in the energy
increment for the step $\Nshell = 4 \rightarrow 5$. Furthermore, we
observe that the energy increment for $\Nshell = 4 \rightarrow 5$ is
approximately the difference between the $\Nshell = 5$ result and the
fully converged result given by NCFC.  It will be valuable to extend
the $\Nshell$ cutoff further in a future effort in order to determine
the full convergence pattern for $^6$Li.  Note that the results
obtained in the FCI $\Nshell = 5$ basis space, with a dimension of 129
million, are very close to the $\Nmax = 6$ results, with a basis space
dimension of less than 0.2 million.  On the other hand, with only 50
Monte Carlo basis states, the MCSM produces a ground state energy that is
within 0.5~MeV of the FCI result at $\Nshell = 5$. With extrapolation,
the ground state energy is within 20~keV of the FCI result with an
extrapolation uncertainty of 83~keV.

\subsection{Convergence of MCSM calculations}

\begin{figure}[tbp]
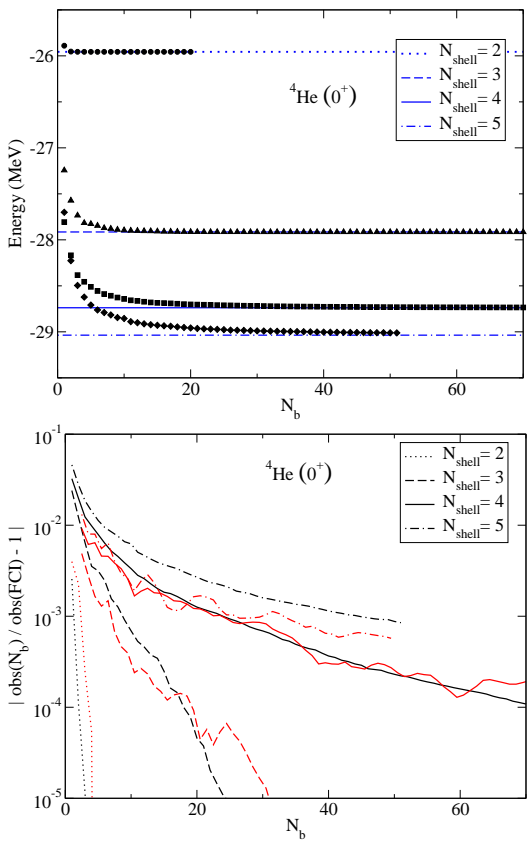

  \includegraphics[width=0.80\columnwidth]{res_4He_Egs.eps}
  \includegraphics[width=0.80\columnwidth]{res_4He_Er.eps}
  \caption{(Color online) 
    The convergence of the MCSM ground state energy of $^4$He to the FCI
    result for several $\Nshell$ values as function of the number 
    of Monte Carlo basis states, $N_b$.  Top: ground state energy;
    Bottom: relative difference between MCSM and FCI calculations of
    the energy (black) and rms matter radius (red).
    \label{Fig:4He}}
\end{figure}
In top panel of Fig.~\ref{Fig:4He} we show the convergence of the
ground state energy of $^4$He as function of the number of the Monte
Carlo basis states, $N_b$.  At every value of $N_b$ the MCSM gives a
variational upper bound for the energy, and as $N_b$
increases, the energy approaches the exact FCI result from above.  In
the bottom panel we show the relative difference between the MCSM and
the FCI result, not only for the energy but also for the rms radius.

Both the ground state energy and the radius of $^4$He converge to
within 1\% of the exact FCI result with less than ten Monte Carlo
basis states, at least up to $\Nshell=5$.  However, in general, as the
number of shells increases, so does the number of Monte Carlo basis
states that is needed in order to achieve a fixed level of accuracy:
at $\Nshell=4$ we need about 20 basis states in order to reach an
agreement of 0.1\%, but at $\Nshell=5$ we need about 50 basis
states to reach the same level of accuracy, as can be seen from the
bottom panel of Fig.~\ref{Fig:4He}.

\begin{figure*}[t]
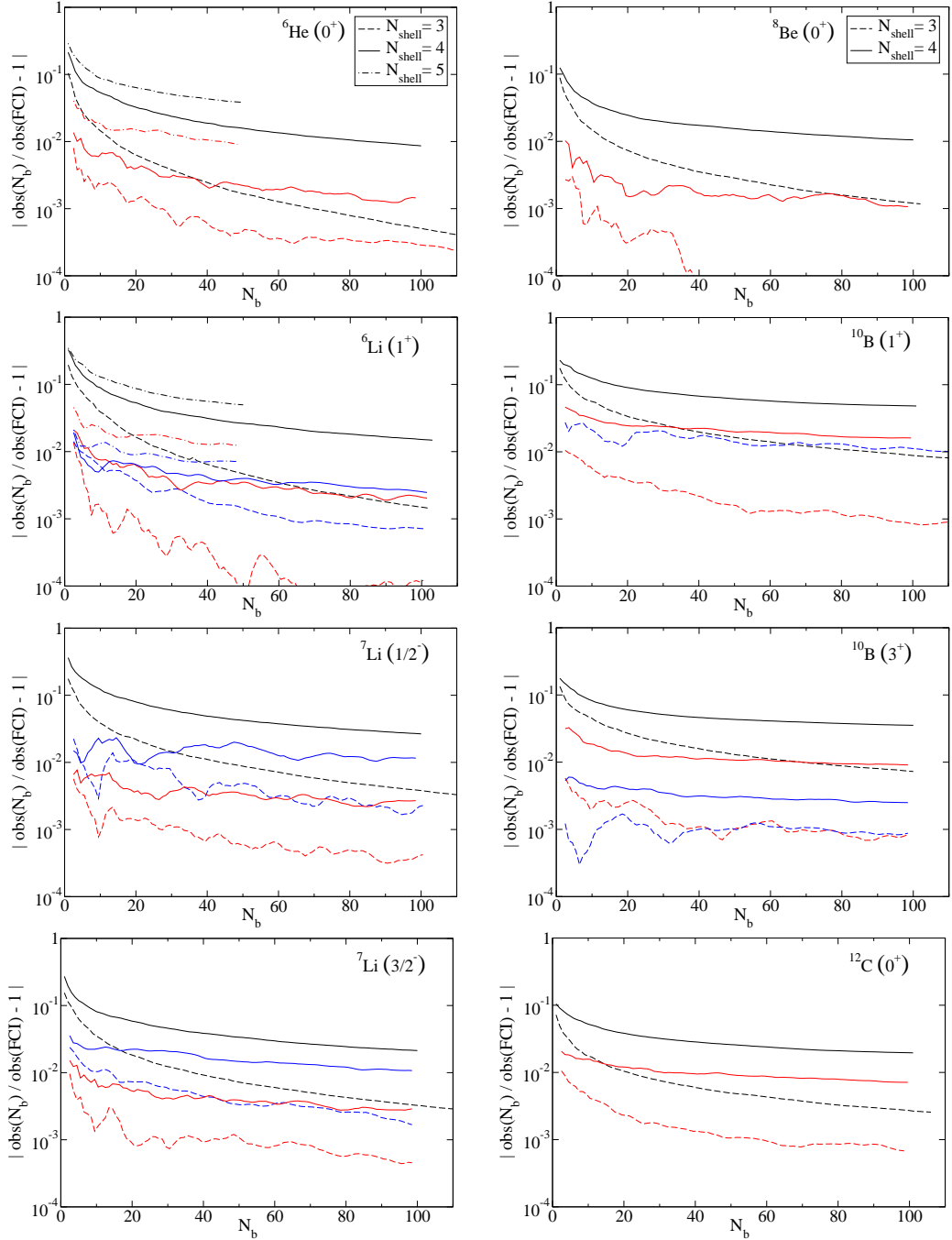

\center{
\includegraphics[width=0.75\columnwidth]{res_6He_Er.eps}\qquad\includegraphics[width=0.75\columnwidth]{res_8Be_Er.eps}
\includegraphics[width=0.75\columnwidth]{res_6Li_Ermu.eps}\qquad\includegraphics[width=0.75\columnwidth]{res_10B1_Ermu.eps}
\includegraphics[width=0.75\columnwidth]{res_7Li1_Ermu.eps}\qquad\includegraphics[width=0.75\columnwidth]{res_10B3_Ermu.eps}
\includegraphics[width=0.75\columnwidth]{res_7Li3_Ermu.eps}\qquad\includegraphics[width=0.75\columnwidth]{res_12C_Er.eps}
}
  \caption{(Color online)
    The relative difference between MCSM and FCI calculations 
    of the energy (black), rms matter radius (red),
    and magnetic moment (blue) for several $\Nshell$ values 
    as function of the number of MC basis states, $N_b$.  
    Top to bottom on the left:
    $^6$He($J^\pi = 0^+$), $^6$Li($1^+$), 
    $^7$Li($\frac{1}{2}^-$), and $^7$Li($\frac{3}{2}^-$);
    Top to bottom on the right:
    $^8$Be($0^+$), $^{10}$B($1^+$), $^{10}$B($3^+$), $^{12}$C($0^+$).
    \label{Fig:reldiffs}}
\end{figure*}

The number of the Monte Carlo basis states needed for a given level of
accuracy depends not only on $\Nshell$, but also on the
number of nucleons $A$, the quantum numbers of the state under consideration,
and the observable, as can be seen from Fig.~\ref{Fig:reldiffs}.

In general, the convergence with $N_b$ to the exact FCI result starts
out very rapidly, but slows down as $N_b$ increases.  The energy
always converges monotonically (at least for the lowest states of a
given spin and parity), because of the variational principle, but
other observables such as the rms radii and the magnetic moments do
not converge monotonically.  On average, however, the difference
between the MCSM results and the FCI results decreases with increasing
$N_b$, as one would expect.

Furthermore, the average convergence rate with increasing $N_b$ 
for different observables of a particular state at fixed $\Nshell$ tends to be the same.
That is, if the MCSM energy converges rapidly to the FCI
energy, then so do the rms radius and magnetic moments of that
state, but if the MCSM energy converges slowly, then the other
observables converge slowly as well, as one can see from
Figs.~\ref{Fig:4He} and \ref{Fig:reldiffs}.  This suggests that 
the wave function obtained with the MCSM converges to the FCI
wave function in a systematic manner that can be measured 
by different observables.

\begin{figure}[t]
\includegraphics[width=0.75\columnwidth]{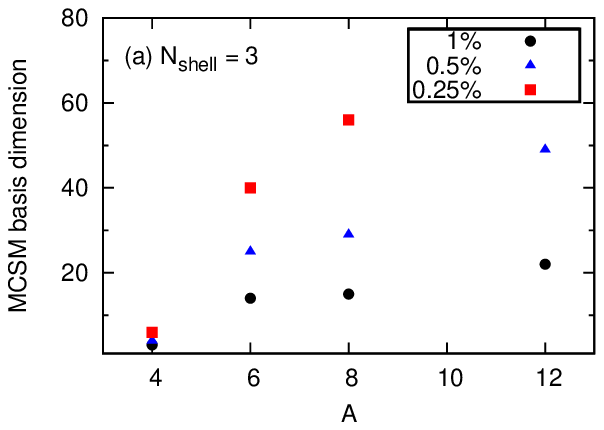}
\includegraphics[width=0.75\columnwidth]{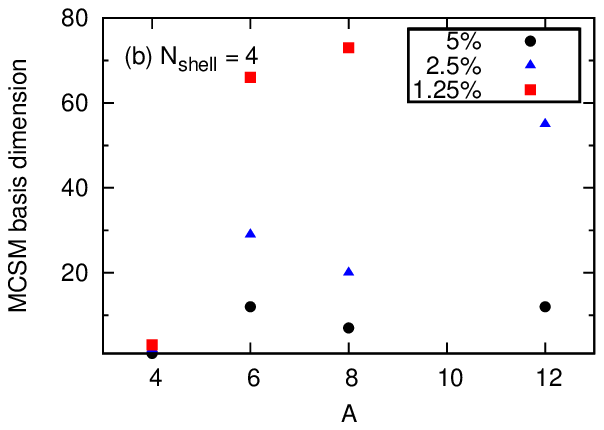}
  \caption{(Color online)
    Number of MC basis states, $N_b$, required for a given accuracy 
    of the MCSM energy for (a) $\Nshell=3$ and (b) $\Nshell=4$ 
    for $^4$He($J^\pi = 0^+$), $^6$He($0^+$), $^8$Be($0^+$), and $^{12}$C($0^+$).
    \label{Fig:numstates}}
\end{figure}
%
In Fig.~\ref{Fig:numstates} we show the number of Monte Carlo basis
states $N_b$ that are needed in order to achieve a specified level of accuracy
for the energy as function of $A$ for the
four nuclei under consideration that have a $0^+$ ground state.
Clearly, the convergence for $^4$He is much faster than for any of the
other nuclei, and its convergence rate is (unfortunately) not a good
indicator for the convergence rate that can be expected for heavier
nuclei.  For the other three nuclei we see that at $\Nshell=3$ a
doubling of the MCSM basis dimension leads to a reduction of the
difference with the FCI results by (approximately) a factor of two.
However, at $\Nshell=4$ one needs to more than double the MCSM basis
dimension in order to improve the accuracy by a factor of two.

Naively, one might expect that the number of Monte Carlo states needed
for a given level of accuracy increases with $A$ (and with $\Nshell$)
proportional to the dimension of the underlying FCI basis space.
However, in practice it turns out that the number of required Monte
Carlo states increases much slower with $A$ than the FCI dimension.
Note that as $A$ increases, the number of pairwise correlations grows
as $A^2$ and one might expect to require a similar increase in Monte
Carlo basis states in order to achieve a given level of accuracy with
strong $NN$ interactions.  Hence one could expect a much more modest
increase in the number of required Monte Carlo states for a given
accuracy than the dramatic growth with $A$ of the dimension of the
underlying FCI basis at fixed $\Nshell$, see Table~\ref{Table:Nshell}.
Indeed, for $\Nshell=3$ this dependence seems to be roughly between
linear and quadratic in $A$, though for $\Nshell=4$ the trend is not
very clear.  Also, so far we have only looked at $p$-shell nuclei, and
it is as of yet unclear how convergence behaves in the $sd$ shell.

\subsection{Extrapolation to FCI}

To obtain the converged energy at fixed $\Nshell$ we
extrapolate the MCSM results by using the energy variance, which is a
new ingredient of the MCSM approach~\cite{ref6}.  The energy variance
$\Delta E_2$ is defined as 
\begin{equation}
  \Delta E_2 = \langle \Psi | H^2 | \Psi \rangle 
     - \left(\langle \Psi | H | \Psi \rangle\right)^2.
\end{equation}
For an eigenstate of $H$, the energy variance is zero, but if $\Psi$
is not an exact eigenstate of $H$ the energy variance is larger than
zero.  As we increase $N_b$, the number of Monte Carlo states in the
MCSM calculations, we get a better and better approximation of the
(lowest) eigenstates of $H$.  Therefore, $\Delta E_2$ approaches zero
from above as $N_b$ increases.  We use this to obtain an estimate of
the exact FCI answer.

\begin{figure}[t]
\includegraphics[width=0.75\columnwidth]{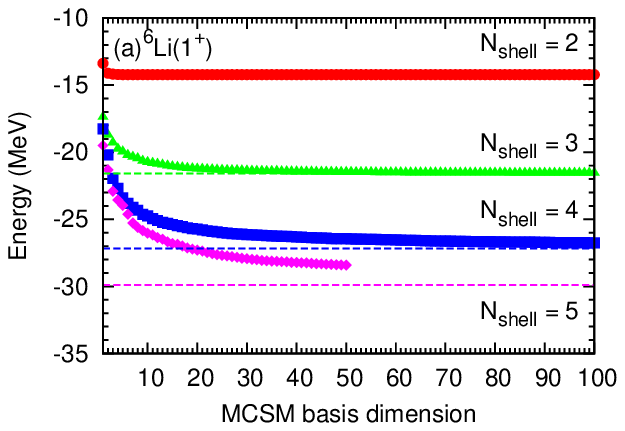}
\includegraphics[width=0.80\columnwidth]{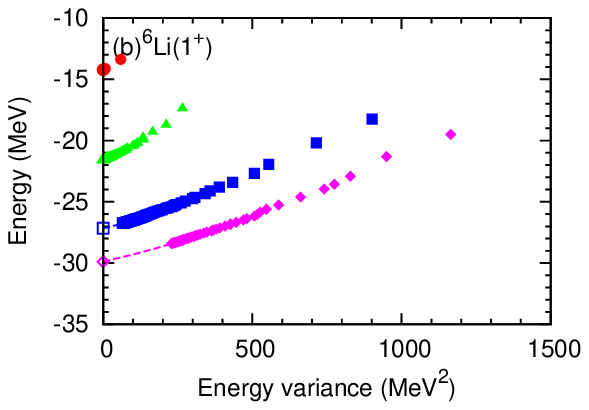}
  \caption{(Color online)
    The convergence of the MCSM ground state energy of $^6$Li to the
    FCI result; circles (red), triangles (green), squares (blue), and
    diamonds (purple) indicate the results at $\Nshell =$ $2$, $3$,
    $4$ and $5$, respectively.  (a) Ground state energy as function
    of the number of Monte Carlo basis states, $N_b$; (b) Energy
    variance $\Delta E_2$ and extrapolation to zero energy variance.
    \label{Fig:6LiE}}
\end{figure}

We plot the MCSM results for the ground state energy of $^6$Li
at different values of $N_b$ as a function of
the evaluated energy variance $\Delta E_2$, see Fig.~\ref{Fig:6LiE}.
For $\Nshell=2$ and $3$ (red and green symbols), the MCSM energy
converges rapidly to the FCI result (top panel), and the energy
variance goes to zero (bottom panel).  For $\Nshell=4$ and $5$ (blue
and purple symbols), the energy variance does decrease with increasing
$N_b$, but does not reach zero in our calculations.  For comparison,
the open symbols at $\Delta E_2=0$ are the results of our (exact) FCI
calculations.

The behavior of energy as function of the energy variance is monotonic
and can be extrapolated to zero energy variance (which corresponds to
the exact energy) by quadratic fitting functions as was done in
Ref.~\cite{ref6},  
\begin{equation}
  E(\Delta E_2) = c_0 + c_1 \Delta E_2 + c_2 (\Delta E_2)^2  
\end{equation}
with the fit parameters, $c_0$, $c_1$, and $c_2$. 
Here, $c_0$ gives the exact energy, $E(\Delta E_2 = 0)$. 
Indeed, the extrapolations for $\Nshell=4$ and $5$
reproduce the exact FCI results to within a few tens of keV, well
within the numerical uncertainty in the extrapolation.
The numerical uncertainty for the extrapolation is estimated based on
the uncertainties $\delta c_i$ in each of the three fit parameters
$c_i$ of the quadratic fit.  We treat these three uncertainties as
independent, and combine them at the MCSM result with minimum energy
variance, $x = \min(\Delta E_2)$, to produce an overall estimate of
the extrapolation uncertainty
\begin{equation}
 \delta E =
\sqrt{
\bigg(\frac{\delta E}{\delta c_0}\bigg|_x \, \delta c_0\bigg)^2
+ \bigg(\frac{\delta E}{\delta c_1}\bigg|_x \, \delta c_1\bigg)^2
+ \bigg(\frac{\delta E}{\delta c_2}\bigg|_x \, \delta c_2\bigg)^2 }.
\end{equation}

The FCI and the MCSM results for the energies with and without
the energy variance extrapolation are all summarized in
Table~\ref{Binding_energies}.  Note that we also quote the estimated
uncertainty from extrapolation in Table~\ref{Binding_energies}.

We use a similar extrapolation for the rms matter radii and, if
possible, also for the magnetic dipole and electric quadrupole
moments.  However, the approach of these observables to the exact FCI
result is generally not monotonic, and therefore not as easy to
extrapolate.  In practice we use a linear extrapolation for these
observables, and apply the extrapolation only if the energy variance
plot appears to be linear.  The detailed dependence of both the energy
and the other observables on the energy variance is presented in the
Appendix~\ref{Appendix}.

\subsection{Point-particle rms radii }

\begin{figure}[tbp]
\includegraphics[width=0.80\columnwidth]{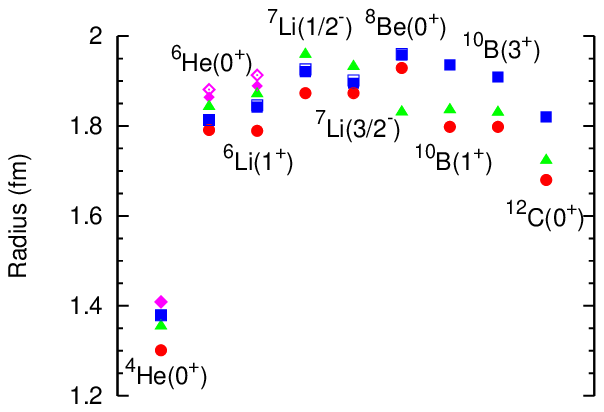}
\caption{ (Color online) 
  Comparisons of the point-particle rms matter radii between the MCSM
  and FCI.  Circles (red), triangles (green), squares (blue) and
  diamonds (purple) indicate the results at $\Nshell =$ $2$, $3$, $4$
  and $5$, respectively; solid (open) symbols stand for the MCSM (FCI)
  results.  The MCSM results are those after extrapolation as described in the
  text. Note that the $^{10}$B and $^{12}$C results at $\Nshell = 4$
  were obtained only within the MCSM.
  \label{Mass_RMS}}
\end{figure}

\begin{table}[tbp]
    \begin{tabular}{lllll}
      \hline
      \hline
& & \multicolumn{3}{c}{$\sqrt{\langle r^2 \rangle}$ (fm)} \\
\cline{3-5}
Nuclei & Method & $\Nshell=3$ & $\Nshell=4$ & $\Nshell=5$ \\
      \hline
        $^4$He ($0^+$)   $\quad$ & MCSM $\quad$ & 1.355 $\qquad$ & 1.379 $\qquad$ & 1.409  $\qquad$ \\
                         \quad &extrp&       &       & 1.410(1) \\ 
                         \quad &FCI &  1.355 & 1.379 & 1.410 \\
        $^6$He ($0^+$)   &MCSM&  1.843 & 1.811 & 1.864 \\
                         &extrp& 1.843(1) & 1.813(1) &    \\
                         &FCI &  1.843 & 1.813 & 1.881 \\
        $^6$Li ($1^+$)   &MCSM&  1.871 & 1.842 & 1.889 \\
                         &extrp& 1.871(1) & 1.846(1) &    \\
                         &FCI &  1.871 & 1.846 & 1.913 \\
        $^7$Li ($1/2^-$) &MCSM&  1.958 & 1.921 &       \\
                         &extrp& 1.959(1) & 1.925(1) &    \\
                         &FCI &  1.959 & 1.926 &       \\
        $^7$Li ($3/2^-$) &MCSM&  1.931 & 1.895 &       \\
                         &extrp& 1.932(1) & 1.900(1) &    \\
                         &FCI &  1.932 & 1.901 &       \\
        $^8$Be ($0^+$)   &MCSM&  1.831 & 1.958 &       \\
                         &extrp& 1.831(1) & 1.960(1) &    \\
                         &FCI &  1.831 & 1.960 &       \\
      $^{10}$B ($1^+$)   &MCSM&  1.834 & 1.936 &       \\
                         &extrp& 1.836(1) & 1.967(2) &    \\
                         &FCI &  1.836 &       &       \\
      $^{10}$B ($3^+$)   &MCSM&  1.829 & 1.909 &       \\
                         &extrp& 1.830(1) & 1.926(1) &    \\
                         &FCI &  1.830 &       &       \\
      $^{12}$C ($0^+$)   &MCSM&  1.722 & 1.820 &       \\
                         &extrp& 1.723(1) & 1.833(1) &    \\
                         &FCI &  1.723 &       &       \\
      \hline
      \hline
    \end{tabular}
\caption{
  Point-particle rms matter radii (in fm) 
  evaluated relative to the nuclear c.m.  The entries labeled ``MCSM"
  indicate the MCSM results before the energy variance extrapolation,
  while the those of the ``extrp'' line denote the MCSM results after
  the extrapolations.  See Table~\ref{Binding_energies} for $N_b$ and
  $\hbar\omega$ values.
  \label{RMS_radii}}
\end{table}

We present the point-nucleon rms matter radii in Fig.~\ref{Mass_RMS}
and Table~\ref{RMS_radii} calculated with the wave functions of the
MCSM and FCI methods. For this comparison, we evaluate the rms radius
of the internal degrees of freedom --- that is we use the radius
operator that depends only on the coordinates with respect to the c.m.
of the system.  Thus, although the nuclear wave functions contain
mixtures of various components of c.m. motion, the use of the internal
coordinates for the radial operator will provide a more accurate rms
radius for eventual comparison with experiment. In addition, at the
present level of benchmark effort this is sufficient to compare
results between these approaches.  As mentioned above, the exact
separation of the c.m. motion from the internal motion is a nontrivial
challenge for the MCSM and FCI approaches while that separation may be
assured in the NCFC approach by use of a constraint on the c.m. motion.

The MCSM results in Fig.~\ref{Mass_RMS} and those in Table \ref{RMS_radii} 
labelled ``extrp" are
obtained by extrapolation with first-order polynomials using their dependence on
the energy variance (see the Appendix for more details).  We find the
differences between the extrapolated MCSM and FCI rms matter radii 
to be less than
0.1\%, and within the estimate of the extrapolation uncertainty.  As a
consequence, the open symbols for FCI lie nearly on top of the solid
symbols for the extrapolated MCSM so that they are not separately visible in
Fig.~\ref{Mass_RMS}.  However, the MCSM results for $^6$He and $^6$Li
in $\Nshell=5$ with only 50 Monte Carlo basis states are not
sufficient for an extrapolation to the exact FCI result; more Monte
Carlo states are needed for a reliable extrapolation for these cases.
Note that the rms results for $^{10}$B and $^{12}$C at $\Nshell = 4$
were obtained only within the MCSM approach.

\subsection{Dipole and quadrupole moments}

\begin{figure}[t]
\includegraphics[width=0.80\columnwidth]{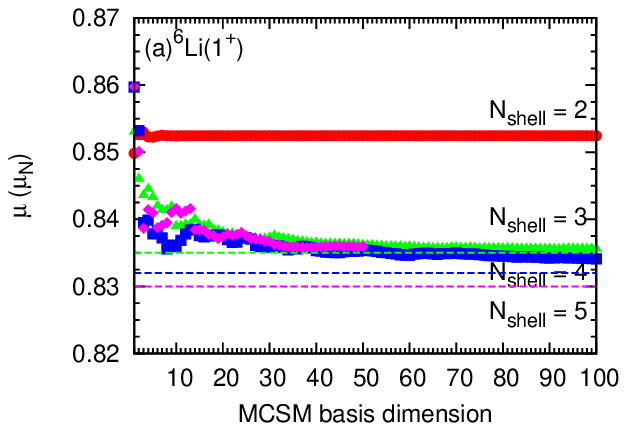}
\includegraphics[width=0.85\columnwidth]{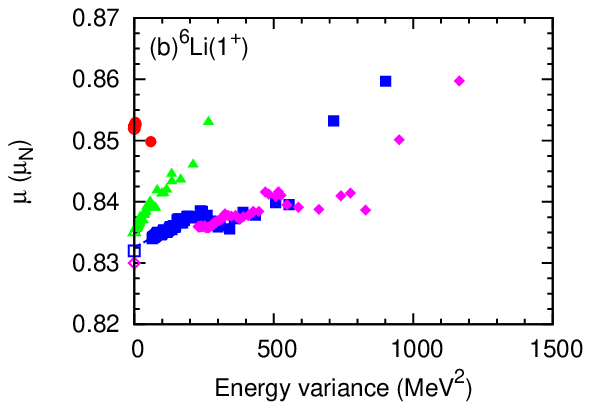}
  \caption{(Color online)
    The convergence of the MCSM magnetic moment of $^6$Li to the FCI
    result; circles (red), triangles (green), squares (blue) and
    diamonds (purple) indicate the results at $\Nshell =$ $2$, $3$,
    $4$ and $5$, respectively.  (a) Magnetic moment as function of
    the number of Monte Carlo basis states, $N_b$; (b) Energy
    variance and extrapolation to zero energy variance.  For
    comparison, the dashed lines (top) and open symbols at $\Delta
    E_2 = 0$ (bottom) are the results of our (exact) FCI calculations.
    \label{Fig:6Limu}}
\end{figure}
%
In Fig.~\ref{Fig:6Limu} we plot the MCSM results for the magnetic
moment of $^6$Li as function of $N_b$ (top) and as function of the
evaluated energy variance $\Delta E_2$.  For $\Nshell=2$ and $3$ (red
and green symbols), the MCSM results converge rapidly to the FCI
result (top panel), and the energy variance goes to zero (bottom
panel).  For $\Nshell=4$ (blue symbols), the MCSM results do seem to
converge to the FCI results, and with a linear extrapolation on the
energy variance we get good agreement with the FCI results.  However,
just as for the rms radius, the MCSM results for $\mu$ at $\Nshell=5$
with only 50 Monte Carlo basis states are not sufficient for an
extrapolation to the exact FCI result; more Monte Carlo states are
needed for a reliable extrapolation for this case.

\begin{table}[tbp]
    \begin{tabular}{lllll}
      \hline
      \hline
 & & \multicolumn{3}{c}{$\mu$ ($\mu_N$)} \\
\cline{3-5}
 Nuclei & Method & $\Nshell=3$ & $\Nshell=4$ & $\Nshell=5$ \\
      \hline
        $^6$Li ($1^+$)   $\quad$ &MCSM $\quad$&    0.836 $\qquad$ &  0.834 $\qquad$ & 0.836 $\qquad$ \\
                         &extrp&   0.835(1) &  0.833(1) &    \\
                         &FCI &    0.835 &  0.832 & 0.830 \\
        $^7$Li ($1/2^-$) &MCSM&    -0.842 & -0.816 &       \\
                         &extrp&   -0.840(1) & -0.806(2) &    \\
                         &FCI &  -0.840 & -0.807 &       \\
        $^7$Li ($3/2^-$) &MCSM&   3.061 &  3.025 &       \\
                         &extrp&  3.057(1) &  2.995(2) &    \\
                         &FCI &   3.056 &  2.993 &       \\
      $^{10}$B ($1^+$)   &MCSM&   0.503 &  0.533 &       \\
                         &extrp&  0.508(1) &        &       \\
                         &FCI &   0.509 &        &       \\
      $^{10}$B ($3^+$)   &MCSM&   1.820 &  1.814 &       \\
                         &extrp&  1.818(1) &  1.819(1) &     \\
                         &FCI &   1.818 &        &       \\
      \hline
  & & \multicolumn{3}{c}{$Q$ ($e$fm$^2$)} \\
      \hline
        $^6$Li ($1^+$)   &MCSM&   -0.259 & -0.282 & -0.276 \\
                         &extrp&  -0.259(1) & -0.285(1) &     \\
                         &FCI &   -0.259 & -0.285 & -0.302 \\
        $^7$Li ($3/2^-$) &MCSM&   -1.766 & -2.006 &        \\
                         &extrp&  -1.750(1) & -1.958(3) &      \\
                         &FCI &   -1.750 & -1.940 &        \\
      $^{10}$B ($1^+$)   &MCSM&   -1.712 & -2.417 &        \\
                         &extrp& -1.703(2) & -2.436(8) &     \\
                         &FCI &  -1.698 &        &        \\
      $^{10}$B ($3^+$)   &MCSM&    3.532 &  5.222 &        \\
                         &extrp&  3.503(1) &  5.250(11) &      \\
                         &FCI &    3.503 &        &        \\
      \hline
      \hline
    \end{tabular}
    \caption{Dipole (top) and quadrupole moments
      (bottom) calculated using the wave functions obtained within the
      MCSM and FCI methods.  The entries of the MCSM indicate the MCSM
      results before the energy variance extrapolation, while the
      those of the ``extrp'' line denote the MCSM results after the
      extrapolations.  See Table~\ref{Binding_energies} for $N_b$ and
      $\hbar\omega$ values.
      \label{moments_table}}
\end{table}
We summarize our comparison of MCSM and FCI results for the magnetic
dipole moments and electric quadrupole moments in
Table~\ref{moments_table}, both with and without the extrapolations of
the MCSM results.  We use a linear extrapolation using the energy
variance, see the Appendix for more details.  As mentioned above, we
cannot perform a reliable extrapolation from the MCSM results to the
exact FCI result for the dipole and quadrupole moments of $^6$Li at
$\Nshell=5$ with only 50 Monte Carlo states.  The only other case for
which we could not perform a reliable extrapolation is the magnetic
moment of the (lowest) $1^+$ state of $^{10}$B.  This is likely to be
related to the fact that there are two $1^+$ states relatively close
to each other (experimentally their energies differ by about
1.5 MeV).

We again find the differences between the MCSM and the exact FCI
results to be small, typically 1\% or less with 100 Monte Carlo
states, both for the magnetic dipole moments and for the electric
quadrupole moments.  The exception is the quadrupole moment of the
ground state of $^6$Li: at $\Nshell=5$, the difference between the
MCSM and FCI calculations is almost 10\% with 50 Monte Carlo basis
states.  Note however that this quadrupole moment is exceptionally
small in magnitude: although the relative difference between the NCSM
and the FCI result is significantly larger than for most other
observables we have looked at, the absolute difference is 
rather small.

A linear extrapolation using the energy variance brings the MCSM
results even closer to the FCI results, see the Appendix for more
details.  However, the results after 50 Monte Carlo states for $^6$Li
at $\Nshell=5$ are not sufficiently close to the FCI result to do such
an extrapolation.

\begin{figure}[tbp]
\includegraphics[width=0.80\columnwidth]{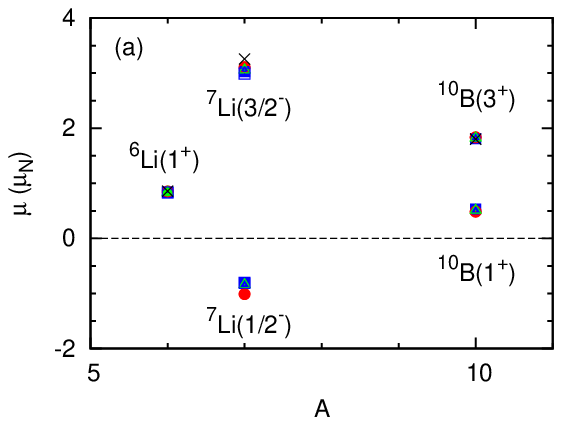}
\includegraphics[width=0.80\columnwidth]{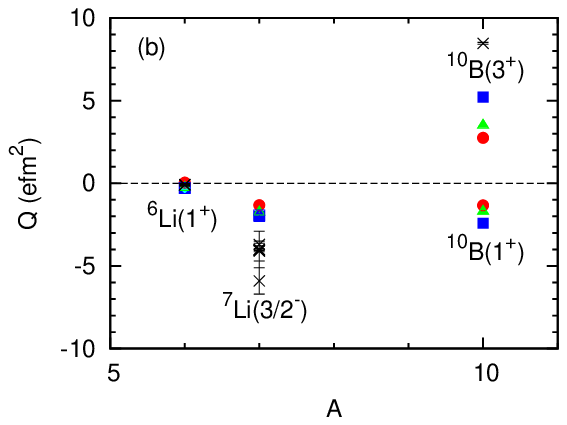}
\caption{ (Color online) 
  Comparisons between the extrapolated MCSM and FCI results for 
  (a) the magnetic dipole moments, and (b) electric quadrupole moments.  
  The conventions for the symbols are same as in
  Fig.~\ref{Mass_RMS}; crosses indicate the experimental values 
  for the ground states from Ref.~\cite{exp_moments}.
  \label{moments_figure}}
\end{figure}
The magnetic dipole moments, top panel of Fig.~\ref{moments_figure},
depend only very weakly on the basis space truncation parameters
$\Nshell$ and $\hbar\omega$ --- much weaker than the quadrupole
moments, bottom panel of Fig.~\ref{moments_figure}, and than the rms
radii, Fig.~\ref{Mass_RMS}.  The differences are less than $2\%$ and
are not visible on this scale.  Furthermore, the dipole moments are in
very good agreement with the NCFC results, which means that they are
converged to within a few percent with respect to the basis space
truncation.  Our results with JISP16 are also in good agreement with
the available data for the ground states.

On the other hand the quadrupole moments do depend significantly on
the basis space truncation parameters $\Nshell$ and $\hbar\omega$, as
can be seen from the bottom panel of Fig.~\ref{moments_figure} and
from Table~\ref{moments_table}.  This is not surprising, given the
dependence of the rms radius on the truncation parameters, and given
the fact that the quadrupole moment receives significant contributions
from the asymptotic tail of the wave function, which is not very well
represented in a HO basis.  One needs to include (much) higher HO
shells in order to build up a realistic tail for the wave functions.
Nevertheless, our results are in qualitative agreement with the
available data: small and negative for the ground state of $^6$Li,
large and negative for the ground state of $^7$Li, large and positive
for the ground state of $^{10}$B.

\section{Summary and Outlook
\label{Sec_5}}

We have performed benchmark calculations of the energies,
point-particle rms matter radii, and electromagnetic moments for nine
states in light nuclei ranging from $^4$He to $^{12}$C.  Where
possible, we have solved for these properties using the FCI, MCSM and
NCFC approaches.  The energies and the point-particle rms
matter radii calculated by MCSM were extrapolated as a function of
energy variance.  All results are found to be consistent with each
other to within quoted uncertainties when they could be quantified.
Where we could not obtain quantified uncertainties, the results were
found to differ typically by a few percent among the available methods
with very few exceptions. 
The MCSM and FCI results are very close to each
other with small differences (of a few percent in most cases) arising
mainly from the limited number of MCSM basis sampled stochastically
for diagonalization and from MCSM energy variance extrapolation
uncertainties.  We include selected NCFC results in order to gauge the
increases in basis spaces needed to better approach the fully
converged results (basis space cutoff independence) in future efforts.

Since the MCSM computational effort scales more favorably with
increasing basis space and increasing nucleon number, we expect that
the MCSM will further develop into a powerful tool for {\it ab initio}
nuclear theory.  To reach this goal, we will need to expand the basis
space, treat the role of c.m. motion and include the Coulomb interaction
as well as $NNN$ interactions.  These challenges will be addressed in
future efforts.

\begin{acknowledgments}
This work was supported in part 
by the SPIRE Field 5 from MEXT, Japan.  
We also acknowledge Grants-in-Aid for Young Scientists
(Nos.~20740127 and 21740204), for Scientific Research
(Nos.~20244022 and 23244049), 
and for Scientific Research on Innovative Areas
(No.~20105003) from JSPS, and the CNS-RIKEN joint project for
large-scale nuclear structure calculations.  This work was also
supported in part by the US DOE Grants No.~DE-FC02-07ER41457,
DE-FC02-09ER41582 (UNEDF SciDAC Collaboration), and DE-FG02-87ER40371,
and through JUSTIPEN under grant no.~DE-FG02-06ER41407.  A part of
the MCSM calculations was performed on the T2K Open Supercomputer at
the University of Tokyo and University of Tsukuba, and the BX900
Supercomputer at JAEA. 
Computational resources for the FCI and NCFC calculations were 
provided by the National Energy Research Supercomputer Center (NERSC), 
which is supported by the Office of Science of the U.S. Department of 
Energy under Contract No. DE-AC02-05CH11231, and by the Oak Ridge 
Leadership Computing Facility at the Oak Ridge National Laboratory, 
which is supported by the Office of Science of the U.S. Department of 
Energy under Contract No. DE-AC05-00OR22725.

\end{acknowledgments}

\appendix*
\section{Extrapolation of MCSM results
\label{Appendix}}

With increasing Monte Carlo basis dimension $N_b$, the MCSM results
converge to the FCI results.  In order to obtain an estimate of that
exact FCI answer, we extrapolate the energy and other observables
evaluated by MCSM using the energy variance.  That is, the MCSM
results are plotted as a function of the evaluated energy variance 
\begin{equation}
  \Delta E_2 = \langle \Psi | H^2 | \Psi \rangle 
     - \left(\langle \Psi | H | \Psi \rangle\right)^2
\end{equation}
and then extrapolated to zero variance as we show below.  We also
investigate the uncertainties of this extrapolation and report those
uncertainties in Tables~\ref{Binding_energies}--\ref{moments_table}.

\begin{figure*}[H]
\center{
\includegraphics[width=0.85\columnwidth]{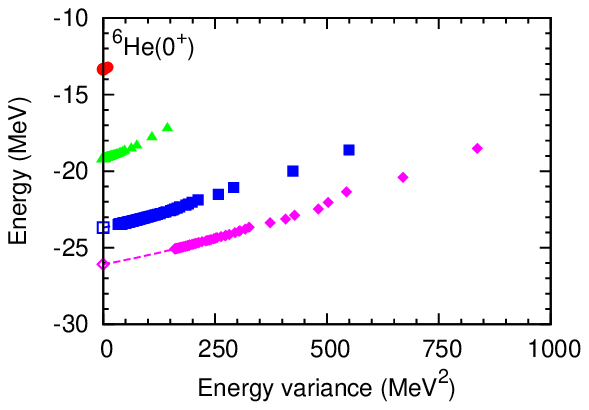}\qquad\includegraphics[width=0.85\columnwidth]{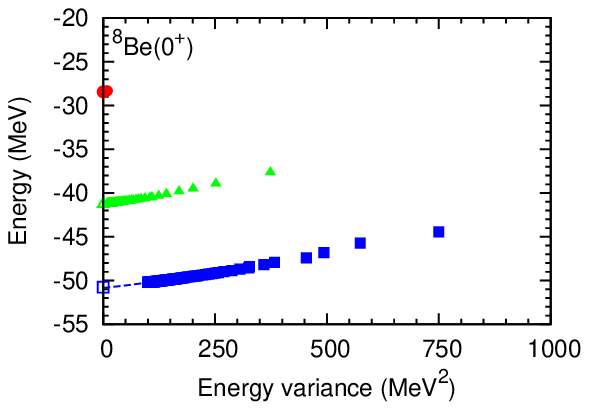}
\includegraphics[width=0.85\columnwidth]{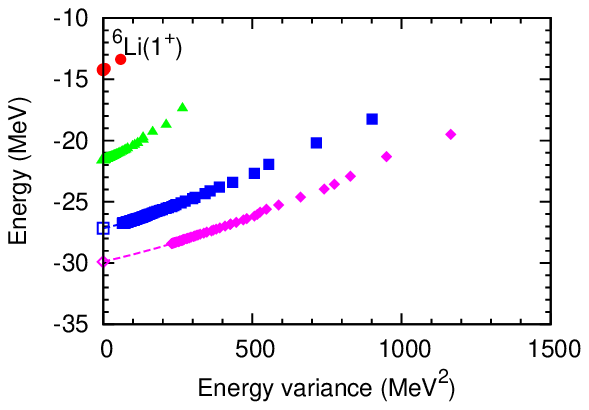}\qquad\includegraphics[width=0.85\columnwidth]{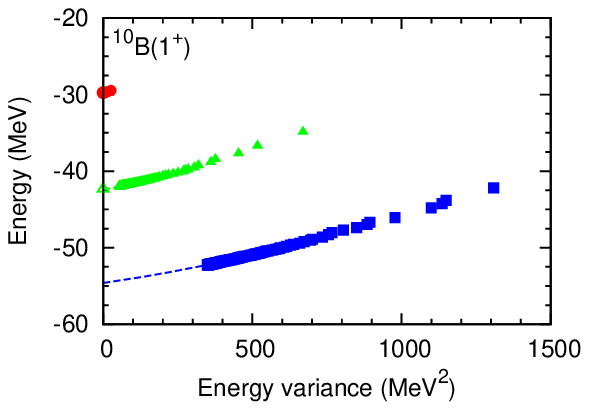}
\includegraphics[width=0.85\columnwidth]{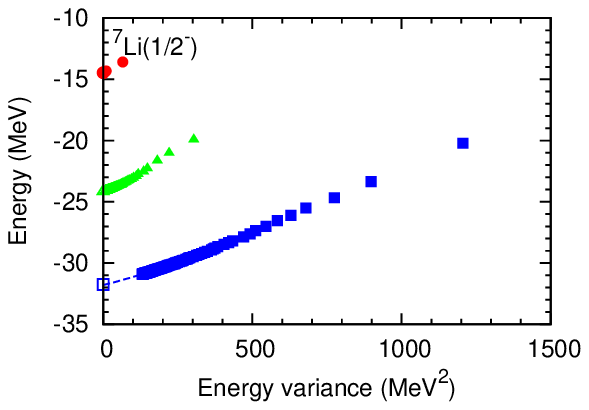}\qquad\includegraphics[width=0.85\columnwidth]{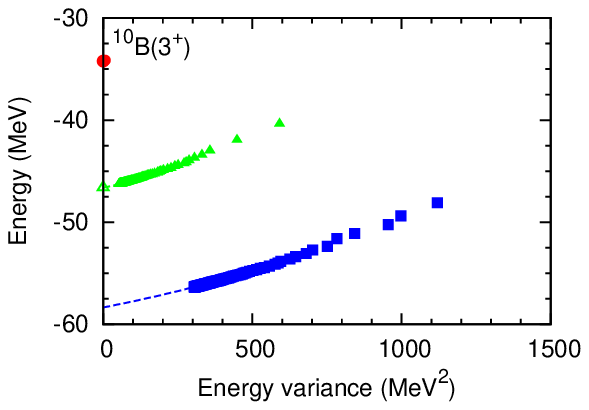}
\includegraphics[width=0.85\columnwidth]{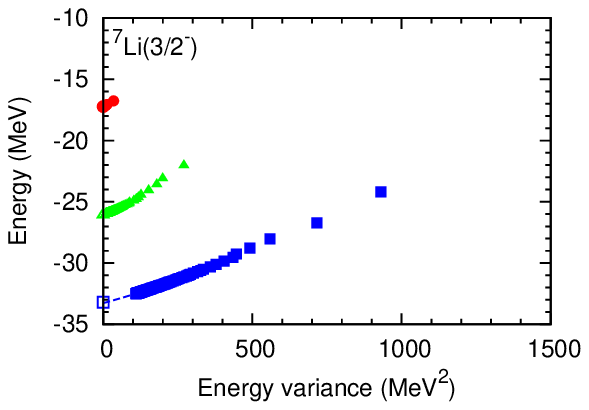}\qquad\includegraphics[width=0.85\columnwidth]{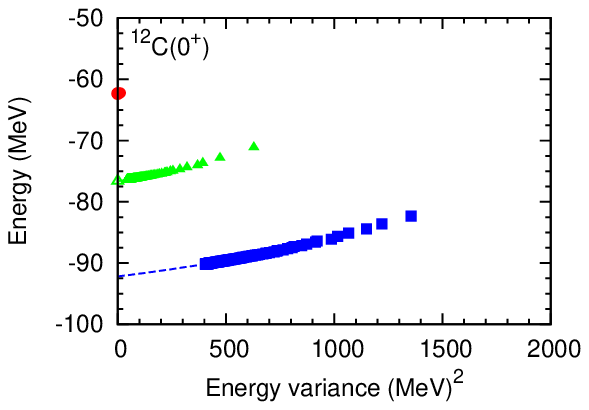}
}
  \caption{(Color online) 
    The energy variance and extrapolation to the FCI result for the
    energies.  Circles (red), triangles (green), squares
    (blue) and diamonds (purple) indicate the results at $\Nshell=2$,
    $3$, $4$ and $5$, respectively; the open symbols at $\Delta E_2=0$
    are the exact FCI energies.  Top to bottom on the left: 
    $^6$He($J^\pi = 0^+$), $^6$Li($1^+$), $^7$Li($\frac{1}{2}^-$), 
    and $^7$Li($\frac{3}{2}^-$); Top to bottom on the right:
    $^8$Be($0^+$), $^{10}$B($1^+$), $^{10}$B($3^+$), and $^{12}$C($0^+$).
  \label{FigA:energies}}
\end{figure*}

Figure \ref{FigA:energies} shows the energies as function of
the energy variance for $^6$He, $^6$Li, $^7$Li, $^8$Be, $^{10}$B, and
$^{12}$C.  For $\Nshell=2$ and $3$ there is no need for any
extrapolation: with 100 Monte Carlo states, there is very good
agreement between the MCSM results and the FCI results.  For
$\Nshell=4$ and $5$ we use a quadratic polynomial fit to extrapolate
$\Delta E_2$ to zero.  We also make an estimate of the numerical
uncertainty in this extrapolation.  These extrapolated MCSM results
are in good agreement with the available FCI results (indicated by the
open symbols at $\Delta E_2=0$).  In Table~\ref{Binding_energies} we
give both the MCSM results, and the extrapolated results with
extrapolation uncertainty.

\begin{figure*}[H]
\center{
\includegraphics[width=0.85\columnwidth]{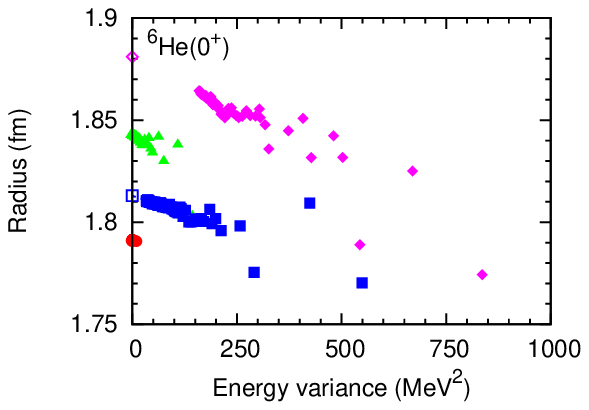}\qquad\includegraphics[width=0.85\columnwidth]{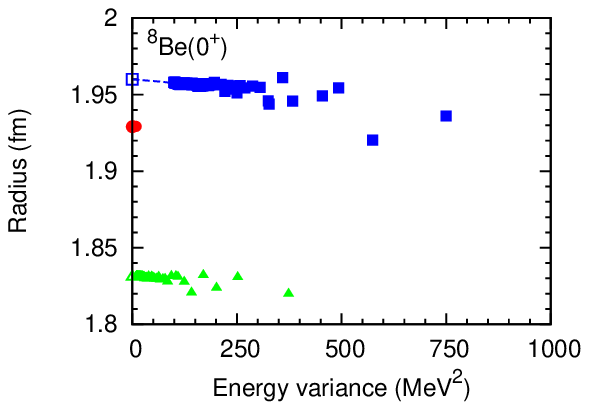}
\includegraphics[width=0.85\columnwidth]{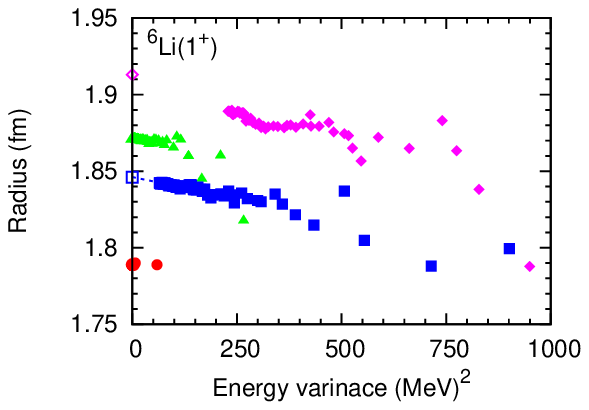}\qquad\includegraphics[width=0.85\columnwidth]{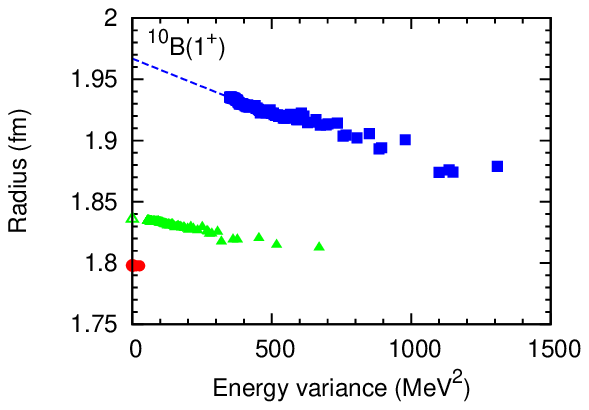}
\includegraphics[width=0.85\columnwidth]{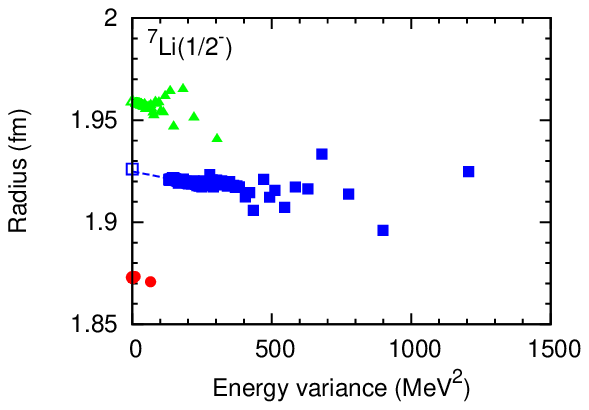}\qquad\includegraphics[width=0.85\columnwidth]{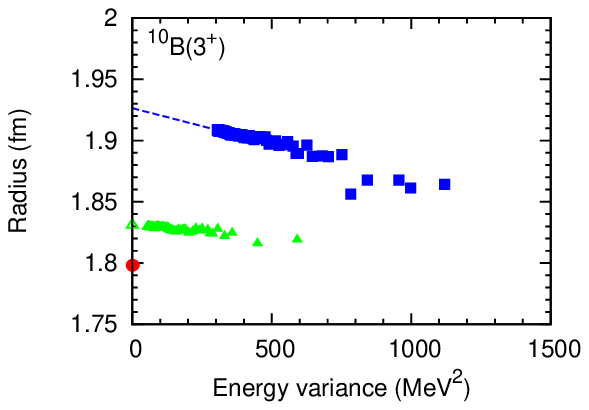}
\includegraphics[width=0.85\columnwidth]{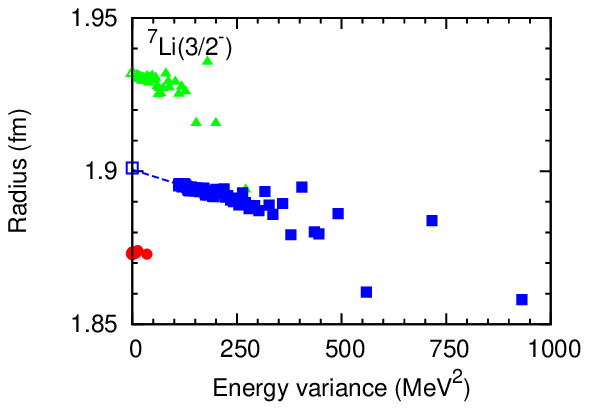}\qquad\includegraphics[width=0.85\columnwidth]{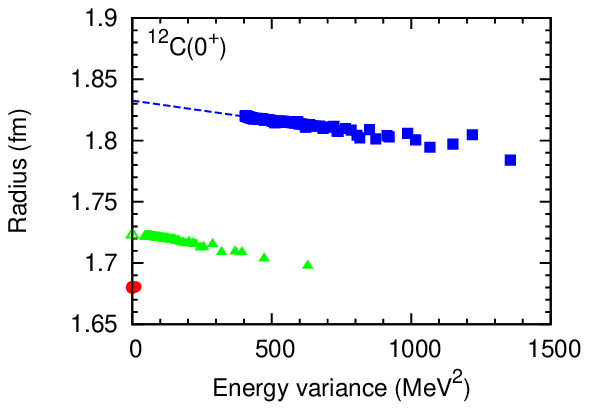}
}
  \caption{(Color online) 
    The energy variance and extrapolation to the FCI result 
    for the rms radii.  Symbols are the same as in
    Fig.~\ref{FigA:energies}. Top to bottom on the left: 
    $^6$He($J^\pi = 0^+$), $^6$Li($1^+$), $^7$Li($\frac{1}{2}^-$), 
    and $^7$Li($\frac{3}{2}^-$); Top to bottom on the right:
    $^8$Be($0^+$), $^{10}$B($1^+$), $^{10}$B($3^+$), and $^{12}$C($0^+$).
    \label{FigA:radii}}
\end{figure*}
\begin{figure*}[H]
\center{
\includegraphics[width=0.80\columnwidth]{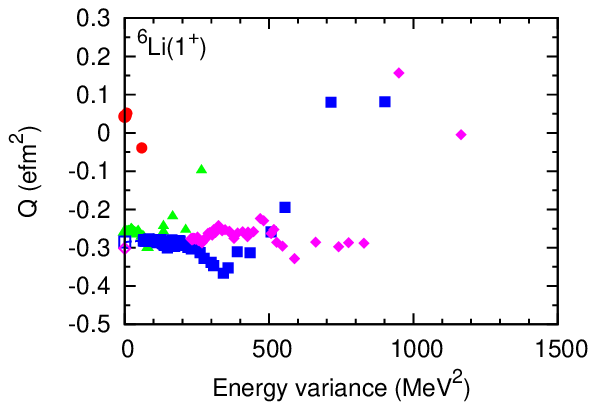}\qquad\includegraphics[width=0.80\columnwidth]{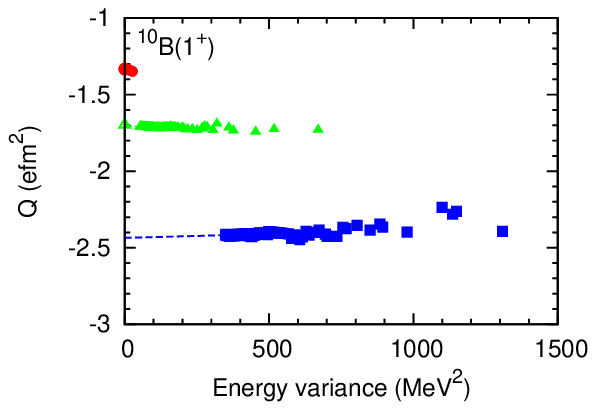}
\includegraphics[width=0.80\columnwidth]{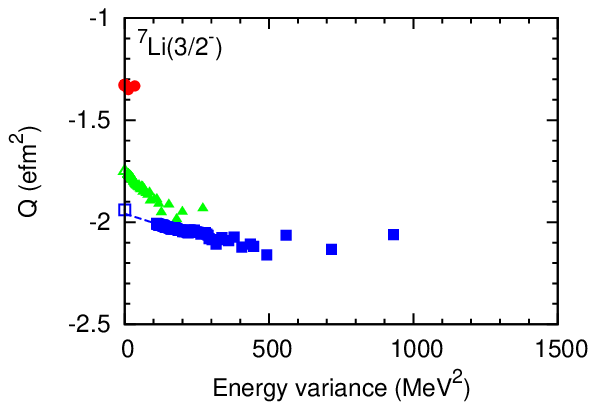}\qquad\includegraphics[width=0.80\columnwidth]{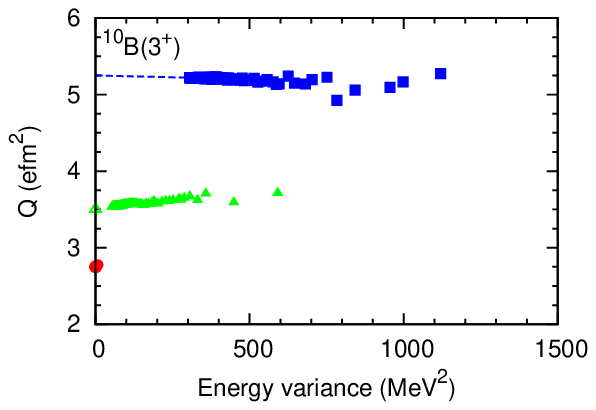}
}
  \caption{(Color online) 
    The energy variance and extrapolation to the FCI result 
    for the quadrupole moments.  Symbols are the same as in
    Fig.~\ref{FigA:energies}.  Top and bottom on the left: 
    $^6$Li($1^+$) and $^7$Li($\frac{3}{2}^-$); 
    Top and bottom on the right:
    $^{10}$B($1^+$) and $^{10}$B($3^+$).
    \label{FigA:quad}}
\end{figure*}
\begin{figure*}[H]
\center{
\includegraphics[width=0.80\columnwidth]{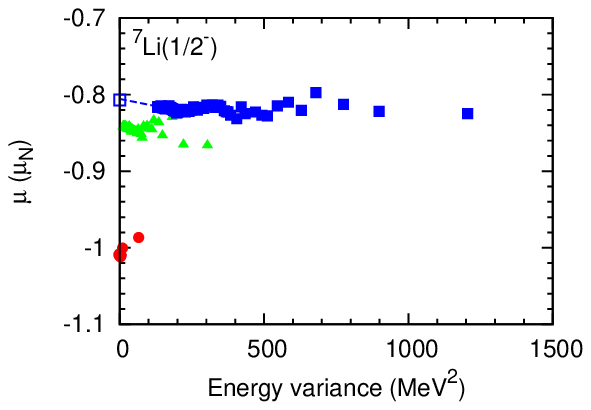}\qquad\includegraphics[width=0.80\columnwidth]{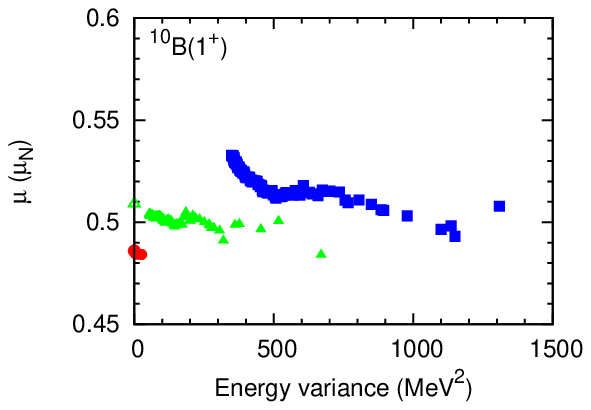}
\includegraphics[width=0.80\columnwidth]{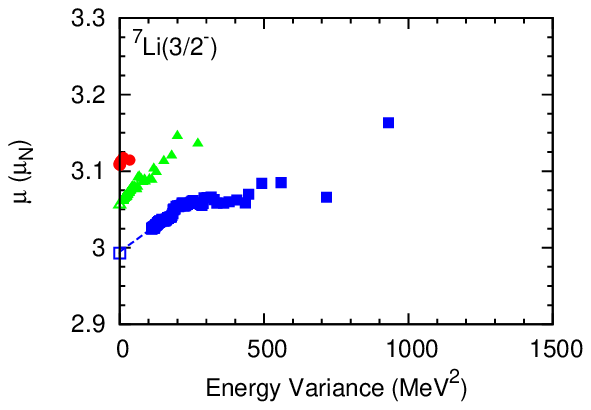}\qquad\includegraphics[width=0.80\columnwidth]{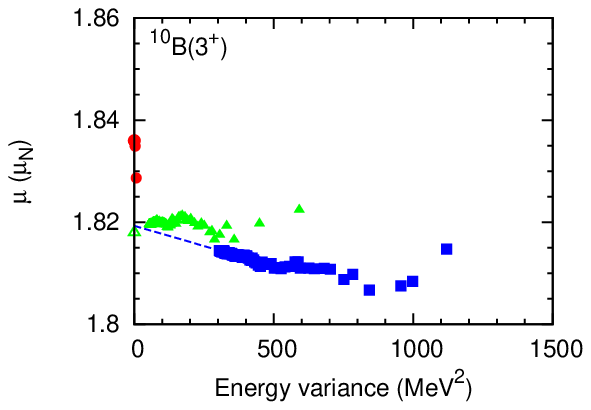}
}
  \caption{(Color online) 
    The energy variance and extrapolation to the FCI result 
    for the dipole moments.  Symbols are the same as in
    Fig.~\ref{FigA:energies}.  Top and bottom on the left:
    $^7$Li($\frac{1}{2}^-$) and $^7$Li($\frac{3}{2}^-$); Top and
    bottom on the right: $^{10}$B($1^+$) and $^{10}$B($3^+$).
    \label{FigA:magn}}
\end{figure*}

Figure \ref{FigA:radii} shows the rms matter radii as function of the
energy variance for $^6$He, $^6$Li, $^7$Li, $^8$Be, $^{10}$B, and
$^{12}$C.  For $\Nshell=2$ and $3$ there is no need for any
extrapolation: with 100 Monte Carlo states, there is very good
agreement between the MCSM results and the FCI results.  For
$\Nshell=4$ we use a linear fit to extrapolate $\Delta E_2$ to zero.
We also make an estimate of the numerical uncertainty in this
extrapolation.  These extrapolated MCSM results are in good agreement
with the available FCI results (indicated by the open symbols at
$\Delta E_2=0$).  In Table~\ref{RMS_radii} we give both the MCSM
results, and the extrapolated results with extrapolation uncertainty.

Unfortunately, 50 Monte Carlo states is not sufficient to extrapolate
the radii of $^6$Li and $^6$He at $\Nshell=5$: the purple diamonds in
the upper left figures cannot be extrapolated reliably to $\Delta E_2=0$.
This is also the case for the magnetic dipole moment, see
Fig.~\ref{moments_figure}, and the electric quadrupole moment, see
Fig.~\ref{FigA:quad} below.

Figure \ref{FigA:quad} shows the electric quadrupole moments as function
of the energy variance for the states that have $J \ge 1$.  For
$\Nshell=2$ there is no need for any extrapolation. However, both the
$\Nshell=3$ and the $\Nshell=4$ MCSM results with 100 Monte Carlo
states can be improved by a linear extrapolation.  As already
mentioned, 50 Monte Carlo states is not sufficient to extrapolate the
quadrupole moment of $^6$Li at $\Nshell=5$.  

Finally, in Fig.~\ref{FigA:magn} we show the magnetic dipole moments
as function of the energy variance for the $^7$Li and $^{10}$B states
that have $J \ge 1$.  Again, for $\Nshell=2$ there is no need for any
extrapolation. Both the $\Nshell=3$ and the $\Nshell=4$ MCSM
results with 100 Monte Carlo states can be improved by a linear
extrapolation.  However, the dependence of the magnetic moment of the
(lowest) $1^+$ state of $^{10}$B does not seem to converge as the
energy variance decreases.  This is possibly caused by the proximity
of a second $1^+$.

The extrapolated MCSM results for both the magnetic moments and the
quadrupole moments are in good agreement with the available FCI
results (indicated by the open symbols at $\Delta E_2=0$).  In
Table~\ref{moments_table} we give both the MCSM results, and the
extrapolated results with extrapolation uncertainty.

\clearpage


\begin{thebibliography}{99}

\bibitem{GFMC}
S.~C.~Pieper, R.~B.~Wiringa, and J.~Carlson,
Phys. Rev. C {\bf 70}, 054325 (2004); 
K.~M.~Nollett, S.~C.~Pieper, R.~B.~Wiringa, J.~Carlson, and G.~M.~Hale
Phys. Rev. Lett. {\bf 99}, 022502 (2007);
S.~C.~Pieper, in 
{\it Proceedings of the International School of Physics ``Enrico Fermi"}, 
Course CLXIX, edited by A. Covello, F. Iachello and R. A. Ricci 
(Societ Italiana di Fisica, Bologna, 2008),  p.~111,
arXiv:0711.1500 [nucl-th]; reprinted in La Rivista del Nuovo Cimento, {\bf 31}, 709, (2008),
and references therein.

\bibitem{NCSM12} 
P.~Navr{\'a}til, J.~P.~Vary, and B.~R.~Barrett, Phys.~Rev.~Lett.~{\bf 84}, 5728 (2000); 
Phys.~Rev.~C {\bf 62}, 054311 (2000); 
S. Quaglioni and P.~Navr{\'a}til, Phys.~Rev.~Lett.~{\bf 101}, 092501 (2008); 
Phys.~Rev.~C {\bf 79}, 044606 (2009).

\bibitem{CC}
G. ~Hagen, T. ~Papenbrock and M. ~Hjorth-Jensen,
Phys. Rev. Lett. {\bf 104} 182501(2010),
and references therein.

\bibitem{Epelbaum} E. ~Epelbaum, W. ~Gl\"ockle, and Ulf-G. ~Meissner,
Nucl. Phys. A {\bf 637}, 107 (1998); {\bf 671}, 295 (2000).

\bibitem{N3LO} D. ~R. ~Entem and R. ~Machleidt,
Phys. Rev. C {\bf 68}, 041001(R) (2003).

\bibitem{Wiringa:1994wb}
  R.~B.~Wiringa, V.~G.~J.~Stoks and R.~Schiavilla,
  Phys.\ Rev.\  C {\bf 51}, 38 (1995).

\bibitem{Pieper_3NF}
S.~C.~Pieper, V.~R.~Pandharipande, R.~B.~Wiringa, and J.~Carlson,
Phys. Rev. C {\bf 64}, 014001 (2001)

\bibitem{Illinois}
S.~C.~Pieper, AIP Conf. Proc. No.~{\bf 1011}, 143(2008).

\bibitem{Shirokov07} A.~M.~Shirokov, J.~P.~Vary, A.~I.~Mazur and T.~A.~Weber,
                   Phys. Letts. B {\bf 644}, 33 (2007); A.~M.~Shirokov, J.~P.~Vary, A.~I.~Mazur, S.~A. Zaytsev 
                   and T.~A.~Weber, {\it ibid}.~{\bf 621}, 96 (2005); subroutines to generate this interaction
                   in the relative-center-of-mass HO basis are available at nuclear.physics.iastate.edu.

\bibitem{Otuska_MCSM} 
M.~Honma, T.~Mizusaki and T.~Otsuka,
Phys.\ Rev.\ Lett.\  {\bf 75}, 1284 (1995); 
{\bf 77}, 3315 (1996); 
T.~Otsuka, M.~Honma and T.~Mizusaki,
{\it ibid}.~{\bf 81}, 1588 (1998); 
for review and further references, see 
T.~Otsuka, M.~Honma, T.~Mizusaki, N.~Shimizu, and Y.~Utsuno, Prog.~Part.~Nucl.~Phys.~{\bf47}, 319 (2001). 

\bibitem{Maris09_NCFC} 
P.~Maris, J.~P.~Vary, A.~M.~Shirokov, Phys. Rev. C {\bf 79}, 014308 (2009);
P.~Maris, A.~M.~Shirokov and J.~P.~Vary, 
{\it ibid}.~{\bf 81}, 021301(R) (2010);
C.~Cockrell, J.~P.~Vary and P.~Maris, 
{\it ibid}.~{\bf 86}, 034325 (2012).

\bibitem{CC_CM}
G.~Hagen, T.~Papenbrock and D.~J.~Dean,
Phys.~Rev.~Lett. {\bf 103}, 062503 (2009).

\bibitem{Vary92_MFDn} J.~P.~Vary, The
Many Fermion Dynamics Shell Model Code, Iowa State University,
1992 (unpublished); J.~P.~Vary and D.~C.~Zheng, {\it ibid}., 1994  (unpublished);
P.~Sternberg, E.~G.~Ng, C.~Yang, P.~Maris, J. P.~Vary, M.~Sosonkina, and H.~V.~Le, 
in {\it Proceedings of the 2008 ACM/IEEE conference on Supercomputing}
(IEEE Press, Piscataway, NJ, 2008), pp.~15:1--15:12.

\bibitem{Maris-ICCS}
P.~Maris, M.~Sosonkina, J.~P.~Vary, E.~G.~Ng and C.~Yang, 
International Conference on Computer Science, ICCS 2010, 
Procedia Computer Science 1, {\bf 97} (2010).

\bibitem{ref5}
N.~Shimizu, Y.~Utsuno, T.~Abe, and T.~Otsuka, RIKEN Accel.~Prog.~Rep.~{\bf 43}, 46 (2010).

\bibitem{Utsuno:2012vm}
Y.~Utsuno, N.~Shimizu, T.~Otsuka and T.~Abe, 
Comput.~Phys.~Comm.~{\bf 184}, 102 (2013).

\bibitem{ref6}
N.~Shimizu, Y.~Utsuno, T.~Mizusaki, T.~Otsuka, T.~Abe, and M.~Honma, Phys.~Rev.~C{\bf 82}, 061305(R) (2010); 
AIP Conf.~Proc.~No.~{\bf 1355}, 138 (2011).

\bibitem{ref7}
T.~Abe, P.~Maris, T.~Otsuka, N.~Shimizu, Y.~Utsuno, and J.~P.~Vary, AIP Conf.~Proc.~{\bf 1355}, 173 (2011).

\bibitem{Puddu}
  G.~Puddu,
  arXiv:1201.0600 [nucl-th].

\bibitem{Roth}
  R.~Roth,
  Phys.\ Rev.\  C {\bf 79}, 064324 (2009).

\bibitem{Dytrych} 
T.~Dytrych, K.~D.~Sviratcheva, C.~Bahri, J.~P.~Draayer and J.~P.~Vary, 
Phys. Rev. Lett. {\bf 98}, 162503 (2007);
T.~Dytrych, K.~D.~Sviratcheva,  C.~Bahri, J.~P.~Draayer and J.~P.~Vary, 
J. Phys. G. {\bf 35}, 095101 (2008);
{\it ibid}. {\bf 35}, 123101 (2008).

\bibitem{Liu2011}
L.~Liu, T.~Otsuka, N.~Shimizu, Y.~Utsuno and R.~Roth, 
 Phys.\ Rev.\  C {\bf 86}, 014302 (2012).

\bibitem{exp_moments}
N.~J.~Stone, Atomic Data and Nuclear Data Tables 90, 75 (2005).

\end{thebibliography}
\end{document}